\documentclass[useAMS,usenatbib,usegraphicx]{mn2e}
\usepackage{aas_macros}
\usepackage{color}
\usepackage{array}
\usepackage{amsmath,amssymb}

\begin{document}

  \title[time-correlated signals in pulsar timing]{Understanding and analysing
    time-correlated stochastic signals in pulsar timing}
  
  \author[van Haasteren and Levin]{
Rutger~van Haasteren$^{1,2}$\footnotemark, Yuri~Levin$^{3,2}$
  \\
    $^1$Max-Planck-Institut f\"ur Gravitationsphysik (Albert-Einstein-Institut),
      D-30167 Hannover, Germany
  \\
    $^2$Leiden Observatory, Leiden University, P.O. Box 9513, NL-2300 RA
      Leiden, the Netherlands
  \\
    $^3$School of Physics, Monash University, P.O. Box 27, VIC 3800, Australia}
 
  \date{printed \today}

  \maketitle

%-- ABSTRACT -------------------------------------------------------------------
  \begin{abstract}
    Although it is widely understood that pulsar timing observations generally
    contain time-correlated stochastic signals (TCSSs; red timing noise is of
    this type), most data analysis techniques
    that have been developed make an assumption that the stochastic
    uncertainties in the data are uncorrelated, i.e. ``white''. Recent work has
    pointed out that this can introduce severe bias in determination of
    timing-model parameters, and that better analysis methods should be used. This
    paper presents a detailed investigation of timing-model fitting in 
    the presence of TCSSs, and gives closed expressions for the
    post-fit signals in the data. This results in a Bayesian technique to obtain
    timing-model parameter estimates in the presence of TCSSs, as well as
    computationally more efficient expressions of their marginalised posterior
    distribution. A new method to analyse hundreds of mock dataset realisations
    simultaneously without significant computational overhead is presented, as
    well as a statistically rigorous method to check the internal consistency of
    the results.  As a by-product of the analysis, closed expressions of the rms
    introduced by a stochastic background of gravitational-waves in
    timing-residuals are obtained, valid for regularly sampled data. Using $T$
    as the length of the dataset, and $h_c(1\rm{yr}^{-1})$ as the characteristic
    strain, this is:
    $\sigma_{\rm GWB}^2 = h_{c}(1\rm{yr}^{-1})^2
      ( 9\sqrt[3]{2\pi^4}\Gamma(-10/3) / 8008)
      \rm{yr}^{-4/3} T^{10/3}$.
  \end{abstract}
%-- abstract -------------------------------------------------------------------


\footnotetext{Email: vhaasteren@gmail.com}

%-- MAIN TEXT ------------------------------------------------------------------
  \section{Introduction}
    Over the years, pulsar timing has proved to be a useful tool for probing a
    wide range of science. Prime examples include the confirmation
    of the emission of gravitational waves \citep{Taylor1982}, and very accurate
    tests of general relativity \citep{Kramer2006}.
    An overview of techniques used in pulsar timing is given in
    \citet{Lorimer2005}, and a detailed description of current treatment of
    timing data is given in the {\rm Tempo2} papers \citep{Hobbs2006,
    Edwards2006}.  Much of the interesting science results from the fact that
    accurate measurements of the times of arrival (TOAs) of the radio pulses
    allow one to precisely track the pulsar trajectories relative to the Earth,
    and and that the observed TOAs can be accounted for very precisely by
    building a physical model of pulsar trajectory, pulse propagation, and
    pulsar spin evolution in relativistic gravity. Such a model is referred to
    as the {\it timing-model}; mathematically the parameters of the timing-model
    are determined by the $\chi^2$ minimisation for the TOA fit.
    
    The remaining differences between the TOAs and the timing-model
    are called timing-residuals (TRs).  The physics beyond that included in  the
    timing-model is contained in TRs. Of particular interest are the
    time-correlated stochastic signals (TCSSs), examples of which
    include:\newline
    1) the so-called pulsar timing noise, or ``red spin noise'', a generic term
    for random changes in the pulsar rotational rate, thought to be possibly due
    to the random angular-momentum exchange between the normal and superfluid
    components,\newline
    2) the time-dependent influence of the interstellar medium on
    the optical pathlength between the pulsar and the Earth, and\newline
    3) the
    influence of the stochastic background of gravitational waves (GWB) on the
    pulse TOAs.\newline
    For a recent discussion of all of these, see \citet[][]{Cordes2010,
    Shannon2010}.

    All of the above processes feature a red spectrum, and their contributions
    to the TOAs are difficult to disentangle from the variation of the
    deterministic timing-model parameters.  The purpose of this paper is to
    develop a rigorous and efficient procedure for a TOA analysis which includes
    simultaneously the timing-model and a TCSS.  The two
    specific questions which are explicitly addressed are:\newline
    1) How does the timing-model fitting affect the statistical properties of
    TCSS-induced timing-residuals?\newline
    2) Conversely, how does the presence of a TCSS affect the uncertainties
    of the timing-model parameters?\newline
    Our method, as well as our answers to the above questions, are extensively
    tested on mock data.

    The detailed plan of the paper is as follows
    In Section~\ref{sec:timingmodelfits} we develop a formalism that casts
    fitting of the timing-model in the presence of red noise as a non-orthogonal
    projection of the covariance matrix.
    A connection with
    least-squares fitting methods is made \citep[as used by e.g.][hereafter
    CHCMV]{Coles2011}. This leads to understanding of degeneracies between the
    covariance matrix
    and the timing-model.
    We show how to exploit these degeneracies in 
    Section~\ref{sec:exploiting},
    which results in improved
    expressions for the covariance function and the marginalised posterior
    distribution.
    We describe how to obtain estimates for the timing-model parameters in
    Section~\ref{sec:timingmodelanalysis}. In Section~\ref{sec:mocktests} we
    introduce a computationally efficient method to analyse hundreds of mock datasets
    simultaneously, and we describe a powerful test based on the
    Kolmogorov-Smirnov statistic that we use to check that the Bayesian analysis
    method produces consistent results.
    Finally, we
    compare our results with the recently proposed Cholesky method of CHCMV in
    Section~\ref{sec:comparison}.

  \section{Timing-model fits and the covariance function}
  \label{sec:timingmodelfits}
    The observed TOAs of every pulsar contain contributions
    from many deterministic and stochastic processes. The traditional procedure
    ignores any stochastic (except TOA uncertainties) or unknown deterministic
    contributions to the timing-residuals \citep[e.g. the standard weighted
    least-squares fit in
    {\rm Tempo2}, see][]{Hobbs2006}.
    Therefore, some of these get absorbed
    into the timing-model fits, which also alters the estimates of the
    timing-model parameters. In this section we show how to model TCSSs in
    combination with fitting to the timing-model.

    \subsection{The covariance function and least-squares fitting}
      We describe the $n$ TOAs of a single pulsar as an addition of a deterministic
      and a stochastic part:
      \begin{equation}
	\vec{t}^{\rm \, arr} = \vec{t}^{\rm \, det} + \vec{\delta t}^{\rm rgp},
	\label{eq:toasource}
      \end{equation}
      where the $n$ elements of $\vec{t}$ are the observed TOAs,
      $\vec{t}^{\text{det}}$ are the deterministic contributions to the TOAs,
      and $\vec{\delta t}^{\rm rgp}$ are the stochastic
      contributions to the TOAs, which in this work are TCSSs all modelled by a
      random Gaussian process.
      In practice, the pre-fit timing
      residuals are produced with first estimates $\beta_{0i}$ of the $m$ timing-model
      parameters $\beta_i$; this initial guess is usually precise enough so that
      a linear approximation of the timing-model can be used further on
      \citep{Edwards2006}. Namely, it is a good assumption that the remaining
      timing-residuals depend linearly on $\xi_a = \beta_a -
      \beta_{0a}$:
      \begin{equation}
	\vec{\delta t} = \vec{\delta t}^{\rm prf} + M\vec{\xi},
	\label{eq:prefitresidual}
      \end{equation}
      where $\vec{\delta t}$ are the timing-residuals in the linear
      approximation to the timing-model, $\vec{\delta t}^{\rm prf}$ is the
      vector of pre-fit timing-residuals, $\vec{\xi}$ is the vector with
      timing-model parameters, and the $(n\times m)$ matrix $M$ is the so-called
      design matrix \citep[see e.g. \S $15.4$ of][hereafter vHLML]{press1992,
      vanhaasteren2009},
      which describes how the timing-residuals depend on the model
      parameters. Without loss of generality, we assume here and in subsequent
      sections that $M$ has been constructed such that its columns are linearly
      independent. Although the distinction between deterministic and TCSS
      contributions here seems analogous to Equation~(\ref{eq:toasource}), we
      note here that there is significant absorption of TCSSs in
      the fit.

      The deterministic signals in $\vec{t}^{\rm \, det}$ are well
      modelled in standard pulsar timing packages like e.g., {\rm Tempo2}
      \citep{Hobbs2006}.
      We model the TCSS contributions as
      a random Gaussian process with a covariance matrix defined by:
      \begin{equation}
	\langle \delta t_i^{\rm rgp} \delta t_j^{\rm rgp} \rangle = C_{ij},
	\label{eq:correlations}
      \end{equation}
      where the brackets $\langle\dots\rangle$ denote the ensemble average 
      of the random process, and the indices $i$ and $j$ run
      from $1$ to $n$. The covariance matrix is the
      numerical representation of the covariance function, and we assume here
      that it can be parametrised with parameters $\vec{\phi}$. We use the
      following convention for the Wiener-Khinchin theorem to relate the
      covariance function to the spectral density:
      \begin{equation}
	C(\tau) = \int_{0}^{\infty}\! S(f) \cos(\tau f) \, \rm{d}f ,
	\label{eq:covariancefunction}
      \end{equation}
      where $S(f)$ is the spectral density of $\vec{\delta t}^{\rm rgp}$ as a
      function of frequency, and $\tau = 2\pi|t_1 - t_2|$ is the time difference
      between two observations multiplied with $2\pi$.  We emphasise here that,
      we do not
      model the post-fit timing-residuals, but the timing-residuals
      $\vec{\delta t}^{\rm rgp}$ prior to the fitting process.
      
      The
      Bayesian likelihood of the timing-residuals is given by (vHLML):
      \begin{eqnarray}
	\label{eq:likelihood}
	P(\vec{\delta t} | \vec{\xi}, \vec{\phi}) &=& \frac{1}{\sqrt{(2\pi)^{n}\det
	C}} \times \\
	& & \exp\left(\frac{-1}{2} \left(\vec{\delta
	t} - M\vec{\xi}\right)^{T}C^{-1}\left(\vec{\delta t} -
	M\vec{\xi}\right)\right). \nonumber
      \end{eqnarray}
      Provided we know the value of the parameters $\vec{\phi}$ prior to the
      analysis, i.e. provided we know the covariance matrix $C$, we can
      maximise the likelihood with respect to the model parameters. This results
      in the
      generalised least-squares (GLS)
      estimator for the timing-model parameters:
      \begin{eqnarray}
	\vec{\chi_{\xi}} &=& \left(M^{T}C^{-1}M\right)^{-1}M^{T}C^{-1} \vec{\delta t}^{\rm prf} \nonumber \\
	\vec{\delta t}^{\rm pof} &=& \left(\mathbb{I}_{n} -
	M\left(M^{T}C^{-1}M\right)^{-1}M^{T}C^{-1}\right) \vec{\delta t}^{\rm
	prf} \nonumber \\
	&=& \left(\mathbb{I}_{n} - B\right)\vec{\delta t}^{\rm pof} = O\vec{\delta t}^{\rm pof},
	\label{eq:gls}
      \end{eqnarray}
      where $\vec{\chi_{\xi}}$ are the best-fit timing-model parameters,
      $\mathbb{I}_{n}$ is the $n$-dimensional identity matrix, $\vec{\delta
      t}^{\rm pof}$ are the post-fit residuals, and $O$ and $B$ are the matrices
      that represent the ``removal'' of the timing-model.
      
      Using an estimate for the spectral density $S(f)$ or the covariance matrix
      $C$ to improve timing model estimates is not new to pulsar timing.
      Firstly, almost
      three decades ago \citet{Blandford1984} analytically showed what the
      effect of the timing noise spectrum is on the timing model parameter
      estimates\footnote{We thank the anonymous referee for bringing this paper
      to our attention. Some of the results in their work were unknown to us
      when we re-derived them.}. Their methods and conclusions are similar: they
      use an analytically derived orthogonal basis to project out the timing
      model basis vectors (our Section~\ref{sec:analyticalcovariance}), and they
      specifically consider a power-law spectral model for the timing noise.
      However, their results do depend on an estimate for the noise; we advocate
      marginalising over those parameters in this work, since we generally do
      not know the values of $\vec{\phi}$ prior to the analysis.

      Secondly, the Cholesky method of CHCMV uses the GLS estimator of
      Equation~(\ref{eq:gls}), combined with an estimate for the covariance
      matrix $C$.

      Thirdly, \citet{Demorest2012} also use the GLS estimator in their efforts
      to constrain or detect an isotropic stochastic gravitational-wave
      background, where they use a Bayesian approach to estimate the noise
      covariance matrix.

      We advocate to marginalise over the parameters of the covariance matrix,
      but we do conclude in this work that for most of the timing model
      parameters the GLS method works well.

    \subsection{Effect of fitting on the covariance function}
      Irrespective of what technique is used to produce the best-fit timing-model
      parameters and the post-fit timing-residuals, the resulting
      post-fit timing-residuals are not correlated according to
      Equation~(\ref{eq:correlations}). In order to
      compute the exact effect that fitting has on the post-fit correlations,
      we take the following approach. Lets assume that we have timing residual vectors of
      length $n$, defined on an interval $[-T,T]$, and some deterministic
      process defined by $m$ parameters represented by $\vec{\xi}$. The effect of
      the deterministic process on the timing-residuals is given by $M\vec{\xi}$.
      Typically, a fitting procedure removes the contributions of the parameters
      $\vec{\xi}$ to the timing-residuals with respect to some inner product. We
      define the inner product on the vector space of timing-residuals as:
      \begin{equation}
	\left\langle \vec{x}, \vec{y}\right\rangle_{E} = \vec{x}^{T} E^{-1} \vec{y},
	\label{eq:innerproduct}
      \end{equation}
      where $x$ and $y$ are vectors of timing-residuals, $E$ is a positive
      definite symmetric (PDS) matrix, and $\langle\dots , \dots \rangle_{E}$
      indicates an inner product with PDS matrix E. Consider a fitting procedure
      that produces post-fit timing-residuals that satisfy:
      $\langle \vec{x}, M\vec{\xi} \rangle_{E} = 0$. It
      is straightforward to check that this is the weighted least-squares fit
      when $E_{ij} = \delta_{ij}\sigma^{2}_{i}$. When $E$
      is a more general PDS matrix, the fitting process that removes all
      $M\vec{\xi}$ from the timing-residuals is the GLS of
      Equation~(\ref{eq:gls}). For our purposes, the exact form of $E$ is not
      relevant.

      We use the expressions for $O$ and $B$ as in Equation~(\ref{eq:gls}), but
      now with covariance function $E$: $B =
      M\left(M^{T}E^{-1}M\right)^{-1}M^{T}E^{-1}$.  Both $O$ and $B$ are
      non-orthogonal projection matrices:
      $B=B^2$, $B^{T}\neq B$ and likewise for $O$. The correlations in post-fit
      timing-residuals due to a random Gaussian process with covariance matrix
      $C$ is given by \citep[see also][]{Demorest2012}:
      \begin{equation}
	\left\langle \vec{\delta t}_i^{\rm pof} \vec{\delta t}_j^{\rm pof}\right\rangle = \left\langle
	\left(\vec{O\delta t}^{\rm pof}_i\right) \left(O\vec{\delta t}^{\rm pof}\right)_j\right\rangle =
	\left(O C O^{T}\right)_{ij}.
	\label{eq:postfitcovariance}
      \end{equation}
      The correlations in the post-fit timing-residuals are thus given by the
      covariance matrix of the random process $C$, projected with the matrix
      $O$, where $O$ removes any contribution $M\vec{\xi}$ with respect to the
      inner product of Equation~(\ref{eq:innerproduct}). From here onwards, we
      omit the superscript ``pof'' for $\vec{\delta t}$, and by default we
      assume we are dealing with post-fit timing-residuals, with respect to some
      inner-product. 
      
  \section{Exploiting fitting degeneracies}
    \label{sec:exploiting}
    Because the fitting process is effectively a projection of the timing
    residuals, the covariance matrix that describes the post-fit residuals is
    also a projection of the pre-fit covariance matrix. This post-fit covariance
    matrix is therefore singular, and the pre-fit covariance matrix cannot be
    reconstructed from the post-fit covariance matrix alone:
    there is a degeneracy in the processes that could have
    generated a single realisation of post-fit timing-residuals. In this
    section we use this degeneracy to derive closed expressions for the
    post-fit covariance function, and more computationally efficient expressions
    for the marginalised posterior distribution.

    \subsection{The post-fit covariance function}
      In Equation~(\ref{eq:postfitcovariance}) we used the non-orthogonal
      projection matrices $O$ and $B$ to remove any contribution $M\vec{\xi}$ to
      the timing-residuals with respect to the inner product of
      Equation~(\ref{eq:innerproduct}). Now consider two related projection
      matrices:
      \begin{eqnarray}
	D &=& M\left(M^{T}M\right)^{-1}M^{T} \nonumber \\
	W &=& \mathbb{I}_{n} - D.
	\label{eq:orthogonalprojections}
      \end{eqnarray}
      Both $D$ and $W$ are orthogonal projections, which satisfy $W^{2}=W$ and
      $W = W^{T}$, and likewise for $D$. The relation with $B$ and $O$ is
      intuitive: if the covariance matrix $E$ of the inner product of
      Equation~(\ref{eq:innerproduct}) is the identity matrix, then we have $W =
      O$, and $D = B$. These four projection matrices have the following
      interesting properties: $BD = D$ and $DB = B$, and similar expressions for
      $W$ and $O$. From this it follows that:
      \begin{equation}
	\left\langle \vec{\delta t} \vec{\delta t}^{T} \right\rangle = O C O^{T}
	= O W C W^{T} O^{T},
	\label{eq:projectedpostfitcovariance}
      \end{equation}
      where the square matrix $\vec{\delta t} \vec{\delta t}^{T}$ is the dyadic product of two
      vectors. This expression shows that
      the degeneracy in the covariance function of the post-fit timing-residuals
      allows us to equivalently use $WCW^{T}$ instead of $C$. In Section
      \ref{sec:analyticalcovariance} we derive analytical expression for this
      post-fit covariance function, see Equation~(\ref{eq:projectedcovariancefunction}).

    \subsection{A simplified marginalised posterior}
      As shown in vHLML, it is possible to analytically marginalise the
      posterior distribution when a flat prior distribution is assumed for the
      linear parameters
      (in Appendix~\ref{sec:appendixb} we show how to include
      Gaussian priors). The
      marginalised posterior distribution is then equal to the
      likelihood function of Equation~(\ref{eq:likelihood}) integrated
      over the linear
      parameters $\vec{\xi}$:
      \begin{eqnarray}
	\int \! \mathrm{d}^{m}\vec{\xi} P(\vec{\delta t} | \vec{\xi},
	\vec{\phi}) &=&
	\frac{\sqrt{\det \left(M^TC^{-1}M\right)^{-1}}}{\sqrt{(2\pi)^{n-m}\det
	C}} \times \nonumber \\
	& & \exp\left(\frac{-1}{2} \vec{\delta
	t}^{T}C'\vec{\delta t}\right),
	\label{eq:marginalisedlikelihood}
      \end{eqnarray}
      with:
      \begin{equation}
	C' = C^{-1} - C^{-1}M\left(M^{T} C^{-1}M\right)^{-1}M^{T} C^{-1}.
	\label{eq:cprime}
      \end{equation}
      Equation~(\ref{eq:marginalisedlikelihood}) and (\ref{eq:cprime}) are the
      computational bottleneck for the analysis of TCSSs in pulsar timing. These
      equations involve non-trivial operations on large, dense matrices.
      Specifically, they involve one $n^3$ operation for the inversion (or
      Cholesky decomposition of $C$), an $m^3$ operation for the inversion of
      $M^{T}CM$, and a lot of vector-matrix operations that scale as $n^2$. In
      this section we seek to simplify these equations for transparency
      and computational efficiency.

      We re-express the effect of the linear parameters $\vec{\xi}$ on the
      timing-residuals in terms of an orthonormal basis. To this end we
      factorise the matrix $M$ with a singular value decomposition:
      \begin{equation}
	M = U \Sigma V^{*},
	\label{eq:svd}
      \end{equation}
      where $U$ and $V$ are respectively $(n\times n)$ and $(m\times m)$
      orthogonal matrices, and $\Sigma$ is an $(n\times m)$ diagonal matrix. For
      our purposes, the column space of the orthogonal matrix $U$ is important.
      The first $m$ columns of $U$ span the column space of $M$, and the last
      $n-m$ columns of $U$ span the complement. We now construct the matrices
      $F$ and $G$ as follows. $U = \begin{array}{cc}(F & G)\end{array}$,
      where $F$ is the $(n\times m)$ matrix consisting
      of the first $m$ columns of $U$, and $G$ is the $(n\times (n-m))$ matrix
      consisting of the other columns of $U$. The following identities
      hold:
      \begin{eqnarray}
	F^{T}F &=& \mathbb{I}_{m} \nonumber \\
	G^{T}G &=& \mathbb{I}_{n-m} \\
	FF^{T} + GG^{T} &=& D + W = \mathbb{I}_{n}.\nonumber
	\label{eq:orthonormalbases}
      \end{eqnarray}
      Using these expressions, it is now possible to show that the marginalised
      likelihood of Equation~(\ref{eq:marginalisedlikelihood}) is equal
      to\footnote{A useful identity is: $$U^{T}CU =$$
      $\begin{pmatrix}
	G^{T}CG & 0 \\
	F^{T}CG & \mathbb{I}_{n-m}
      \end{pmatrix}
      \begin{pmatrix}
	\mathbb{I}_{m} & \left(G^{T}CG\right)^{-1}G^{T}CF \\
	0 & F^{T}CF - F^{T}CG \left(G^{T}CG\right)^{-1}G^{T}CF 
      \end{pmatrix}$}:
      \begin{eqnarray}
	\label{eq:marginalisedlikelihoodnew}
	\int \! \mathrm{d}^{m}\vec{\xi} P(\vec{\delta t} | \vec{\xi}, \vec{\phi})
	&=& \frac{1}{\sqrt{(2\pi)^{n-m}\det
	\left(G^{T}CG\right)}} \times \\
	& &\exp\left(\frac{-1}{2} \vec{\delta
	t}^{T}G \left(G^{T}CG\right)^{-1}G^{T}
	\vec{\delta t}\right). \nonumber
      \end{eqnarray}
      More intuitively, the marginalised likelihood distribution is the
      likelihood function of an $(n-m)$-dimensional random Gaussian process of
      the data $G^{T}\vec{\delta t}$, with a covariance function $G^{T}CG$. This
      more insightful expression involves two matrix-matrix multiplications, and
      an inversion (we ignore the vector-matrix operations, which are $n^2$
      operations).
      Because $G$ is block-diagonal and only needs to be calculated once,
      Equation~(\ref{eq:marginalisedlikelihoodnew}) can be implemented in such a
      way that the main computational burden is the inversion of $G^{T}CG$.
      Since $G^{T}CG$ is a PDS matrix,
      Equation~(\ref{eq:marginalisedlikelihoodnew}) is best evaluated by using
      the Cholesky decomposition, which also directly gives access to the value
      of the required determinant.

      The theoretically more insightful expression we arrive at in
      Equation~(\ref{eq:marginalisedlikelihoodnew})
      is slightly more efficient than
      Equation~(\ref{eq:marginalisedlikelihood}). Also, the dependence on the
      divergent low-frequency cut-off terms have been removed before the
      inversion (see Section~\ref{sec:lfc}), which improves numerical stability.
      However, the final computation still scales as $n^3$, so it will remain a
      computational bottleneck in this type of analysis.

    \subsection{Analytic post-fit covariance functions}
    \label{sec:analyticalcovariance}
      In the previous section, we have shown that we are allowed to use
      $C^{\rm P}=WCW^{T}$ instead of $C$ in all our equations that describe post-fit
      timing-residuals. We analytically approximate that quantity by using
      the inner product of Equation~(\ref{eq:innerproduct}),
      \begin{equation}
	\left\langle\vec{x}, \vec{y}\right\rangle_{E} = \sum_{i}
	\frac{x\left(t_i\right)y\left(t_i\right)}{\sigma^2} \approx
	\frac{1}{\sigma^{2} \Delta t}\int_{-T}^{T}\! x(t)y(t) \, \rm{d}t,
	\label{eq:analyticapprox}
      \end{equation}
      where we have used $E_{ij}=\delta_{ij}\sigma^{2}$, with $\sigma$ the
      uncertainty of the TOAs, $\Delta t$ is the time
      interval between observations, and $x(t)$ and $y(t)$ are continuous
      functions on the interval $[-T,T]$.
      On the interval $[-T,T]$, we define an orthonormal basis of quadratic
      functions \citep[see][for a similar application]{vanhaasteren2010}:
      $\hat{f}_1(t),\hat{f}_2(t),\hat{f}_3(t)$:
      \begin{eqnarray}
	\hat{f}_1(t) &=& \frac{1}{\sqrt{2}}\sigma\sqrt{\frac{\Delta t}{T}}
	\nonumber \\
	\label{eq:orthonormalbasis}
	\hat{f}_2(t) &=& \sqrt{\frac{3}{2}}\sigma\sqrt{\frac{\Delta
	t}{T}}\frac{t}{T} \\
	\hat{f}_2(t) &=& \sqrt{\frac{45}{8}}\sigma\sqrt{\frac{\Delta
	t}{T}}\left[\left(\frac{t}{T}\right)^{2}-\frac{1}{3}\right]. \nonumber
      \end{eqnarray}
      These basis functions satisfy $\langle \hat{f}_{i}, \hat{f}_{j}\rangle_{E} =
      \delta_{ij}$. The process of fitting for quadratics can now be expressed
      as a projection of the covariance functions in terms of these basis
      functions:
      \begin{equation}
	C^{\rm P}(t_0,t_3) = S(t_0, t_1)C(t_1, t_2)S(t_2, t_3),
	\label{eq:projectedcovariancefunction}
      \end{equation}
      where from here onward, we always take inner-product given by
      Equation~(\ref{eq:analyticapprox}) over the repeated variables $t_1$ and
      $t_2$, and $S(t_k, t_l)$ is given by
      \begin{equation}
	S(t_k, t_l) = \sigma^2\Delta t \delta\left(t_k-t_l\right) -
         \sum_{i=1}^{3}
	 \hat{f}_{i}(t_k)\hat{f}_{i}(t_l),
	\label{eq:operatorS}
      \end{equation}
      with $\delta(x)$ the Dirac delta function. Using this formalism, it is
      possible to analytically derive the projected covariance function $C^{\rm
      P}$ for TCSSs with any spectral density.

    \subsection{Power-law covariance function} \label{sec:lfc}
      Power-law spectra are of particular importance in PTA applications, since
      the stochastic background of gravitational waves is expected to be
      a signal well-described by such a power spectral density
      \citep{Begelman1980, Phinney2001, Jaffe2003, Wyithe2003, Sesana2008}.
      Strictly speaking, a process governed by a power-law spectral density is
      improper, and therefore unphysical. For PTA purposes however, the
      power-law behaviour of the signal is expected to hold within its expected
      frequency band of $0.1$--$10$ yr$^{-1}$.
      To enforce finiteness of the covariance function, it is convenient to
      introduce
      cut-off frequencies in
      Equation~(\ref{eq:covariancefunction}); see vHLML. These  can be
      problematic in practice: the cut-off frequency needs to be low enough for
      the signal to represent a power-law signal, yet it must be high enough for
      its numerical representation not to cause numerical artifacts due to
      limited machine precision.
      In this section we show that by choosing the projection matrix to
      represent a fitting procedure that includes quadratic spindown, the
      dependence of the covariance matrix on the low cut-off frequency is
      explicitly removed.

      We parametrise the spectral density as
      \begin{equation}
	S(f) = A^{2}\left(\frac{1}{1 \rm{yr}^{-1}}\right)\left(\frac{f}{1
	\rm{yr}^{-1}}\right)^{-\gamma},
	\label{eq:powerlawspectraldensity}
      \end{equation}
      with $A$ the amplitude of the signal (units time), and $\gamma$ is the
      spectral index. We require a low frequency cut off
      $f_{L}$ if $\gamma \geq 1$. As shown in vHLML, in that case the covariance
      function is equal to:
      \begin{eqnarray}
	C^{\rm PL}_{ij} &=&
	  A^2\left(\frac{1 \rm{yr}^{-1}}{f_{L}}\right)^{\gamma-1}
	  \left\{ \Gamma(1-\gamma)\sin\left(\frac{\pi\gamma}{2}\right)
	  \left(f_{L}\tau_{ij}\right)^{\gamma-1} \right.\nonumber\\
	& &-\left.\sum_{n=0}^{\infty}\left(-1\right)^{n}
	  \frac{\left(f_{L}\tau_{ij}\right)^{2n}}{(2n)!
	  \left(2n+1-\gamma\right)}\right\} . 
	\label{eq:powerlawcovariancefunction}
      \end{eqnarray}
      where $\tau_{ij}=2\pi|t_{i}-t_{j}|$. In vHLML it is shown that the removal
      of quadratic spindown from the timing-residuals also completely removes
      any dependencies on the low-frequency cut off $f_{L}$. In the
      numerical calculations, we need to choose $f_{L} \ll T^{-1}$ so that
      $C^{\rm PL}$ is PDS, and so that we can neglect the terms with $n \geq 2$ in the
      summation of Equation~(\ref{eq:powerlawcovariancefunction}). In practice,
      the diverging terms dependent on the cut-off frequency can result in
      numerical artifacts due to limited machine precision.

      We apply Equation~(\ref{eq:projectedcovariancefunction}) to the covariance
      function of a power-law signal $C^{\rm PL}$ of
      Equation~(\ref{eq:powerlawcovariancefunction}) to obtain the projected
      covariance function $C^{\rm P}(t_0,t_3)$ for a power-law signal. We
      present the details of this
      calculation and the explicit formulae in Appendix~\ref{sec:appendixa}. One
      of the key results of this calculation is that dependencies on the
      low-frequency cut-off $f_L$ are removed from the infinite summation of
      Equation~(\ref{eq:powerlawcovariancefunction}) up to $n=2$. This ensures
      that the quadratic spindown fits have completely removed the sensitivity
      to $f_L$ up to $\gamma<7$. Although this result had been found
      before
      \citep{Blandford1984}, other authors (e.g. vHLML) have assumed that it was
      only true for $\gamma<5$. After this work appeared as a preprint,
      \citet{Lee2012} also obtained both this result, and
      Equation~(\ref{eq:analyticrms}).

      Now that we have an analytic expression for $C^{\rm P}(t_0,t_3)$ for a
      process with the power-law spectral density of
      Equation~(\ref{eq:powerlawspectraldensity}), it is possible to derive an
      expression for the average rms in the post-fit timing-residuals, valid for
      $1<\gamma<7$:
      \begin{eqnarray}
	\sigma^{2}_{\rm PL} &=& \frac{1}{2T}\int_{-T}^{T}\! \rm{d}t C^{\rm P}(t,t) \nonumber \\
	\label{eq:analyticrms}
	  &=&
	  \frac{3(5-\gamma)(\gamma-3)2^{\gamma}(2\pi)^{\gamma-1}}{\gamma(1+\gamma)
	  (3+\gamma)(5+\gamma)} \times \\
	  & &
	  A^{2}\Gamma(1-\gamma)\sin\left(\frac{\pi\gamma}{2}\right)1\rm{yr}^{1-\gamma}T^{\gamma-1} . \nonumber
      \end{eqnarray}
      From this expression, we can derive an estimate for the rms generated by a
      GWB signal of the form:
      \begin{eqnarray}
	\label{eq:characteristicstrain}
	h_c(f) &=& A_h \left(\frac{f}{1 \rm{yr}^{-1}}\right)^{-2/3} \\
	S(f) &=& \frac{A_{h}^{2}}{12\pi^2}1\rm{yr}^{3}\left(\frac{f}{1 \rm{yr}^{-1}}\right)^{-\frac{13}{3}}, \nonumber
      \end{eqnarray}
      where $A_h$ is the dimensionless amplitude of the GWB characteristic
      strain $h_c$. This results in an estimate for the rms of a GWB:
      \begin{eqnarray}
	\label{eq:gwbrmsestimate}
	\sigma_{\rm GWB}^2 &=& A_{h}^{2}
	\left(\frac{9\sqrt[3]{4}\pi^{\frac{4}{3}}\Gamma\left(-\frac{10}{3}\right)}{1001}\right)
	\rm{yr}^{-\frac{4}{3}} T^{\frac{10}{3}} \nonumber \\
	\sigma_{\rm GWB} &=& 4.35 \times 10^{-9}\left(\frac{A_h}{10^{-15}}\right)
	\left(\frac{T}{\rm yr}\right)^{\frac{5}{3}}.
      \end{eqnarray}
      Note here that $2T$ is the total duration of the experiment.

\begin{figure}
	\includegraphics[width=0.5\textwidth]{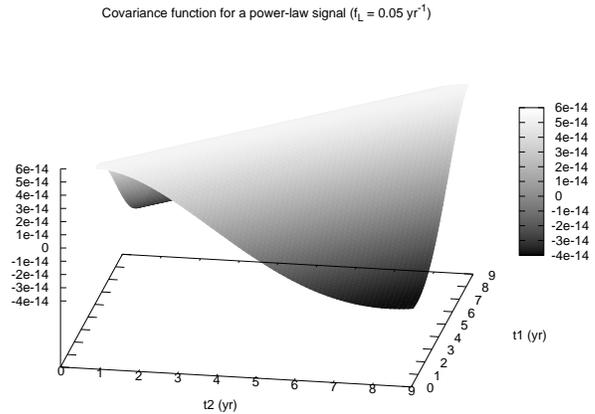}
	\caption{The covariance function $C(t_1,t_2)$ of a power-law spectrum of
	  the form: $S(f) = (A^2 / 1\rm{yr}^{-1}) \times (f/\rm{yr}^{-1})^{-13/3}$, with
	  $A=2.9$ns. For comparison, this is equivalent to a gravitational-wave
	  background with $A_h=10^{-15}$. Here the effect of fitting has been
	  neglected. Notice that the covariance function only depends on the
	  value $t_1 - t_2$.}
	\label{fig:covariance}
      \end{figure}

\begin{figure}
	\includegraphics[width=0.5\textwidth]{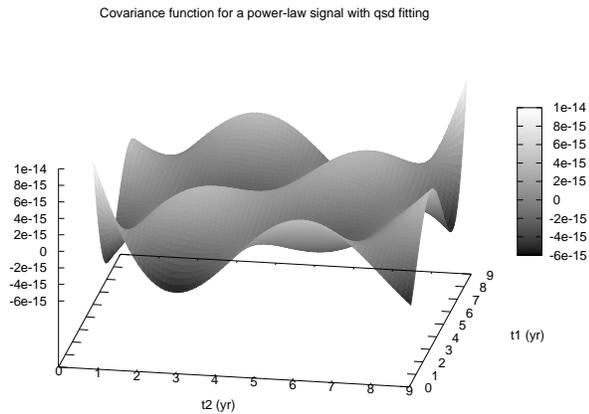}
	\caption{The covariance function $C(t_1,t_2)$ of the same TCSS
	  as in Figure \ref{fig:covariance}.
	  Here the effect of fitting for quadratics has been taken
	  into account analytically with Equation~(\ref{eq:thewholeshebang}) of
	  Appendix~\ref{sec:appendixa}. Notice
	  that, in contrast to Figure \ref{fig:covariance}, the covariance
	  function not only depends on the value $t_1 - t_2$, but on both $t_1$,
	  and $t_2$.}
	\label{fig:covariancean}
      \end{figure}

\begin{figure}
      \includegraphics[width=0.5\textwidth]{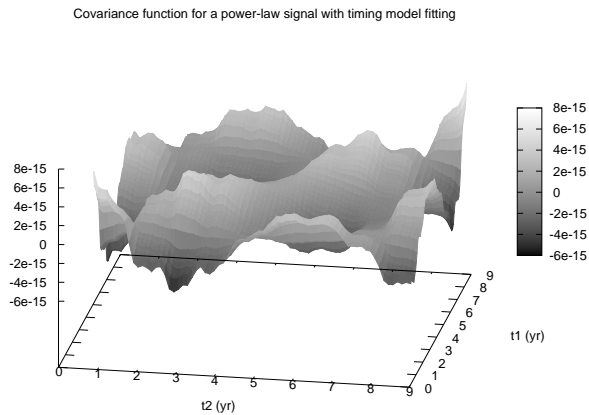}
      \caption{The covariance function $C(t_1,t_2)$ of the same TCSS
	as in Figure \ref{fig:covariance}.
	Here the effect of fitting for the whole timing-model of J$1640$+$2224$
	has been taken into account numerically. 
	Notice
	that, in contrast to Figure \ref{fig:covariance}, the covariance
	function not only depends on the value $t_1 - t_2$, but on both $t_1$,
	and $t_2$. 
	We use the timing-model parameters from the literature
	\citep{Lohmer2005}, and the timing-model as used in {\rm Tempo2}, which
	includes the parameters:
	position (right ascension \& declination), quadratic spindown, proper
	motion (right ascension \& declination), eccentricity, and the projected
	semi-major axis of the binary orbit. The sampling cadence was two weeks.}
      \label{fig:covariancenumall}
    \end{figure}

      As the covariance function is a function of two variables, it is
      instructive to inspect the results of this section visually. First
      consider the pre-fit covariance function $C^{\rm PL}$ of Equation~(\ref{eq:powerlawcovariancefunction}). This is a function of only
      the difference between the two parameters $|t_1-t_2|$, since
      it describes a stationary random process. This is illustrated
      in Figure \ref{fig:covariance}, where lines of equal covariance
      are lines of equal $|t_1-t_2|$. Secondly, we demonstrate what the effect of
      fitting for quadratics is on the covariance function in Figure
      \ref{fig:covariancean}. The symmetry due to the time-stationarity of the
      random process that is present in Figure \ref{fig:covariance} is broken,
      and we see prominent cubical features
      at the edges of the plot.
      Thirdly, we have included the effect of fitting to the
      entire timing-model of an example pulsar in Figure
      \ref{fig:covariancenumall}. This figure is similar to Figure
      \ref{fig:covariancean}, except that there are some small-scale features on
      top of the general structure. This demonstrates that the quadratic
      spindown fitting is the effect that most prominently affects the
      covariance matrix; all other contributions are minor in comparison.

    Because the unweighted least-squares fit is unlikely to be optimal for any
    realistic dataset, it may seem that
    Figures~\ref{fig:covariancean}--\ref{fig:covariancenumall} do not represent a
    post-fit TCSS if a more appropriate fitting procedure is used
    (e.g. the Cholesky method). One may therefore argue that
    Equation~(\ref{eq:gwbrmsestimate}) is not a good measure of the amount of
    detectable GWB signal in the data.  However, the equivalence of $C$ and
    $WCW^{T}$ in Equation~(\ref{eq:projectedpostfitcovariance}) and
    Equation~(\ref{eq:marginalisedlikelihoodnew}) shows that the extra rms in
    the post-fit timing-residuals that may result from an improved fitting
    routine does not increase the sensitivity to a TCSS, since the marginalised
    posterior distribution for the TCSS parameters is always the same. The GWB
    rms quoted in Equation~(\ref{eq:marginalisedlikelihoodnew}) is the part of
    the TCSS that cannot be absorbed in the timing-model: the averaged trace of
    any post-fit covariance function is greater than or equal to this value. The
    quoted GWB rms is therefore a measure of how much ``detectable'' signal is
    in the data.

  \section{Timing-model analysis} \label{sec:timingmodelanalysis}
    When doing an MCMC, we analytically marginalise over the timing-model
    parameters. We would like to retain the information about the timing-model
    parameters, without adding these dimensions to the MCMC. In this section we
    show how to do that efficiently.

    The marginalised posterior of Equation~(\ref{eq:marginalisedlikelihoodnew})
    allows one to numerically marginalise over all the stochastic parameters of
    the model, while analytically marginalising over the timing-model
    parameters with a flat prior. Using that equation it is impossible to obtain
    best estimates for the timing-model parameters.
    Here we show what extra steps
    need to be taken in order to infer the timing-model parameters. We rewrite
    the likelihood of Equation~(\ref{eq:likelihood}) as follows, using the same
    notation as in Equation~(\ref{eq:marginalisedlikelihoodnew}):
    \begin{eqnarray}
      & & P(\vec{\delta t} | \vec{\xi}, \vec{\phi}) = \frac{1}{\sqrt{(2\pi)^{n}\det
      C}} \nonumber \\
      \label{eq:fulllikelihoodnew}
      &\times& \exp\left(\frac{-1}{2}
      \left(\vec{\xi}-\vec{\chi}\right)^{T}\Sigma^{-1}\left(\vec{\xi}-\vec{\chi}\right)\right)
       \\
      &\times&\exp\left(\frac{-1}{2}\vec{\delta
      t}^{T}G\left(G^{T}CG\right)^{-1}G^{T}\vec{\delta t}\right), \nonumber
    \end{eqnarray}
    where $\vec{\chi} = (M^{T}C^{-1}M)^{-1}M^{T}C^{-1}\vec{\delta t}$,
    $\Sigma^{-1} = M^{T}C^{-1}M$,
    and as before, the stochastic parameters are stored in the $l$-dimensional
    vector $\vec{\phi}$. In Appendix~\ref{sec:appendixb} we show how to include
    timing-model parameters with Gaussian priors in a similar manner.
    We are interested in recovering the timing-model parameters
    \begin{equation}
      \vec{\xi} = \sum_{i=1}^{m}\xi_{i}\hat{e_i},
      \label{eq:timingmodelparameters}
    \end{equation}
    where the $\hat{e_i}$ denote the basis vectors of the timing-model
    parameters space.

    The main idea is that in Equation~(\ref{eq:fulllikelihoodnew}), which is
    based on a linear approximation to the timing-model, the likelihood function
    is a multivariable Gaussian with respect to the timing-model parameters. The
    full posterior distribution can be reconstructed without numerically
    exploring the timing-model parameters by:\newline
    1) Constructing a Markov Chain, using the vHLML posterior distribution of
    Equation~(\ref{eq:marginalisedlikelihoodnew}) that faithfully samples the
    stochastic parameters.\newline
    2) Arithmetically averaging the Gaussian timing-model posteriors from each
    point of the chain.\newline
    The full timing-model posterior distribution can then be used to obtain
    marginalised posterior distributions for any combination of timing-model
    parameters. Importantly, as will be demonstrated shortly, all this is done
    without any extra significant computational or memory cost on top of what is
    already used in building a chain in the stochastic parameter space.

    Operationally, as in vHLML, the MCMC is performed using
    Equation~(\ref{eq:marginalisedlikelihoodnew}), where we analytically
    marginalise over the timing-model parameters. However, at each step of
    the Markov Chain, we save the following quantities:
    \begin{itemize}
      \item
	$\vec{\phi}$
      \item
	$P(\vec{\delta t} | \vec{\phi})$
      \item
	$\vec{\chi} = \left(M^{T}C^{-1}M\right)^{-1}M^{T}C^{-1}\vec{\delta t}$
      \item
	$\Sigma^{-1} = M^{T}C^{-1}M$
    \end{itemize}
    This does not require any additional calculations in the MCMC,
    and for each MCMC step the amount of data that
    has to be saved is of the order $m^{2}$, which is not expected to be a
    bottleneck in terms of storage space on modern workstations. We store
    these quantities for each step of the Markov Chain, which has been run in
    the $l$-dimensional parameter space of $\vec{\phi}$, and we thus have enough
    information to fully characterise the ($l$+$m$)-dimensional posterior
    distribution function. Just as in vHLML, the marginalised posterior for the
    stochastic parameters $\vec{\phi}$ can be calculated as usual from the MCMC.

    We assume here that we are interested in calculating the $2$-dimensional
    marginalised posterior as a function two timing-model parameters, say
    $\xi_k$ and $\xi_l$, with $1 \leq k,l \leq m$, but the generalisation to a
    dimensionality other than two is straightforward. The evaluation of the
    $2$-dimensional marginalised posterior consists of numerically integrating
    over the stochastic parameters $\vec{\phi}$ (summing over the MCMC samples),
    and analytically integrating over all but two timing-model parameters along
    the lines of Equation~(\ref{eq:marginalisedlikelihoodnew}).
    Details of this calculation are given in Appendix~\ref{sec:appendixb}, here
    we give the result:
    \begin{equation}
      P\left(\xi_k, \xi_l | \vec{\delta t} \right) = \left\langle
      \frac{\exp\left(\frac{-1}{2}\vec{\Delta
      \xi}^{T}L_{G}\left(L_{G}^{T}\Sigma L_{G}\right)^{-1}L_{G}^{T}\vec{\Delta
      \xi}\right)
      }{\sqrt{\left(2\pi\right)^{2}\det
      \left(L_{G}^{T}\Sigma L_{G}\right)}
      }\right\rangle,
      \label{eq:marginalisedfullposterior}
    \end{equation}
    where we use $\langle\dots\rangle$ to average over all MCMC samples,
    the $(m\times2)$ matrix $L_G = \begin{pmatrix}\hat{e_k} & \hat{e_l}\end{pmatrix}$, with
    $\hat{e_i}$ the $i$-th basis vector for ${\mathbb R}^{m}$, and
    \begin{equation}
      \vec{\Delta \xi} =
      \begin{pmatrix}
	\xi_{k} - \chi_{k} \\
	\xi_{l} - \chi_{l}
      \end{pmatrix}.
      \label{eq:DeltaXi}
    \end{equation}
    Equation~(\ref{eq:marginalisedfullposterior}) allows one to correctly infer
    the parameters of the timing-model, while taking into account the effect of
    red timing noise.

  \section{Tests on an ensemble of mock datasets} \label{sec:mocktests}
    We test the procedures we describe in this work with mock TOAs. The TOAs are
    simulated observations of a pulsar with known timing-model parameters, and
    added noise with known statistical parameters.
    In previous studies, analysing just one dataset with an MCMC was a
    computational challenge \citep[vHLML,][]{vanhaasteren2011}.
    Extensive statistical studies of the behaviour of the data analysis method
    have therefore not been carried out in those studies.
    In this section we show that the Bayesian data analysis method has the
    desired statistical properties by introducing and applying a method with
    which a whole ensemble of mock datasets can be analysed simultaneously without
    much computational overhead.

    \subsection{MCMC and importance sampling}
      Doing a full analysis of a single dataset is a computationally challenging
      task because we have to do non-trivial matrix algebra at each step of the
      MCMC.  This makes a straightforward analysis of a whole ensemble of
      datasets, say $k=1000$ datasets, computationally prohibitive. We seek
      to overcome this problem by analysing a whole ensemble of datasets
      simultaneously when doing only one MCMC simulation. At each sample of the
      chain we efficiently evaluate the likelihood for each dataset, which in
      the end can be used to construct the respective marginalised posterior
      distributions.

      A necessary requirement for doing this, is that
      Equation~(\ref{eq:fulllikelihoodnew}) can be evaluated for each dataset
      without re-doing all the matrix algebra. This is possible if all datasets
      are different realisations of the same process, which is not a restriction
      for the purposes of this work. To ensure realistic simulations, we model
      our mock data after the data for pulsar J$1713$+$0747$ as published in
      \citet{vanhaasteren2011}. This model for our datasets has irregular
      sampling, greatly varying error bars for different TOAs, and an unknown
      jump in the middle of the dataset.  The $n$ TOAs of each dataset are
      generated as perfect realisations of the published timing-model, observed
      at the same MJDs, combined with a TCSS modelled as a
      random Gaussian process with a red spectral density and a flat high
      frequency tail.
      
      We simulate the contributions of the random Gaussian process to the TOAs
      by, for each dataset, appropriately
      transforming a vector of pseudo-random numbers $\vec{\zeta}$ with entries
      drawn from a normal distribution with mean $0$, and width $1$. The
      simulated timing-residuals are then constructed as $\vec{\delta
      t}=L\vec{\zeta}$, with $L$ the lower diagonal Cholesky decomposition of
      the covariance matrix $C$ of Equation~(\ref{eq:covariancefunction}) of the
      random Gaussian process, defined by $C = LL^{T}$. We generate all datasets
      in the ensemble this way, an example
      of which is shown in Figure~\ref{fig:exampleresidualsensemble}. All
      datasets in the ensemble are generated with the same input parameters.

      We evaluate the likelihood function for each dataset $i$, and for each
      MCMC sample $j$, where $i$ runs from $1$ to the number of MCMC samples
      $N$, and $j$ runs from $1$ to $n$.  The samples at which we evaluate the
      likelihood values $L_{ij}$ of all datasets come from an MCMC chain that we
      call the kernel chain. We postpone the details of how we have constructed
      this kernel chain until the next section, for now we assume that we have
      found a suitable kernel, where for each sample we have access to the
      kernel likelihood $L_{0j}$, and the values of the parameters. For a
      canonical MCMC simulation, producing a marginalised posterior distribution
      $p(\theta)$ can be calculated as:
      \begin{eqnarray}
	\label{eq:marginalisedmcmc}
	p( x ) &=& \int\limits_{\theta_{i}=x} \! L( \vec{\theta} ) P_0(
	\vec{\theta} ) \, \mathrm{d}^{m-1}\theta \\ \nonumber
	&\approx& \langle 1 \rangle_{\theta_{i}=x} ,
      \end{eqnarray}
      where $p(\theta_{i})$ is the marginalised posterior, $\theta_{i}$ is the
      $i$-th component of the $m$-dimensional parameter vector $\vec{\theta}$, $L(\vec{\theta})$
      is the likelihood function, $P_0(\vec{\theta})$ is the prior
      distribution, and $\langle\dots\rangle_{\theta_{i}=x}$
      indicates an ensemble average over the MCMC samples over all samples with
      $i$-th parameter equal to $\theta_i$. This expression assumes that the
      samples in the MCMC are sampled with a probability proportional to
      $L(\vec{\theta}) P_{0}(\vec{\theta})$. In our case however, the MCMC
      samples are taken with a probability proportional to the kernel likelihood
      $L_0(\vec{\theta})$. We adjust Equation~(\ref{eq:marginalisedmcmc}) to
      suit this new situation:
      \begin{eqnarray}
	\int\limits_{\theta_{i}=x} \! L( \vec{\theta} ) P_0(
	\vec{\theta} ) \, \mathrm{d}^{m-1}\theta &=&
	\int \! L_{0}( \vec{\theta} ) \frac{L( \vec{\theta} )
	P_0(\vec{\theta} )}{L_{0}( \vec{\theta} )} \,
	\mathrm{d}^{m-1}\theta \nonumber \\
	&\approx& \left\langle \frac{L( \vec{\theta} )
	P_0(\vec{\theta} )}{L_{0}( \vec{\theta} )}
	\right\rangle_{\theta_{i}=x}.
	\label{eq:marginalisedmcmcimportance}
      \end{eqnarray}
      This approach where one samples not from the true distribution, but from a
      distribution similar to the true distribution (the kernel distribution) is
      called importance sampling \citep{Newman1999}, where our samples are
      re-used from the MCMC on the kernel.

    \subsection{Choosing a suitable kernel}
      With Equation~(\ref{eq:marginalisedmcmcimportance}), we can efficiently
      produce marginalised posterior distributions for many datasets at a time.
      Provided there are enough samples in the MCMC, this expression is valid
      for any kernel likelihood function $L_{0}(\vec{\theta})$. However, with
      importance sampling, the efficiency of the MCMC is highly dependent on the
      choice of the kernel likelihood function, with it only being practical if
      the kernel likelihood function is similar to the likelihood functions
      of the datasets. Our datasets satisfy that condition, because they are all
      realisations of the same processes. We therefore take the following
      approach to the construction of a suitable kernel dataset, which is then
      used to form the kernel likelihood function.\newline
      1) We produce a realisation of data, which we call the kernel dataset, in an
      identical manner to how we produced the $k$ datasets.\newline
      2) We randomly delete $4/5$ of the data points in the kernel dataset to ensure
      that the kernel dataset has a broader posterior distribution for all
      parameters than the mock datasets.\newline
      3) The likelihood function that belongs to the kernel dataset is the
      kernel likelihood.\newline
      4) We make sure that for all the parameters that vary during the MCMC,
      that the true value of each parameter is inside the $1$-$\sigma$ region of
      the kernel likelihood. If not, we discard this chain, and start at $1$)
      again to form a new kernel dataset. This step makes sure that our
      particular realisation is not a so-called 'outlier' for our model
      parameters.\newline
      By constructing a kernel likelihood like this, we are ensured that our
      kernel likelihood distribution covers all the high probability density
      (HPD) regions of all the likelihood functions, which allows for faster
      convergence of Equation~(\ref{eq:marginalisedmcmcimportance}). We note
      though that convergence is ensured for \emph{any} kernel dataset.

    \subsection{Statistical properties of the ensemble}
    \label{sec:statisticalproperties}
\begin{figure}
	\includegraphics[width=0.5\textwidth]{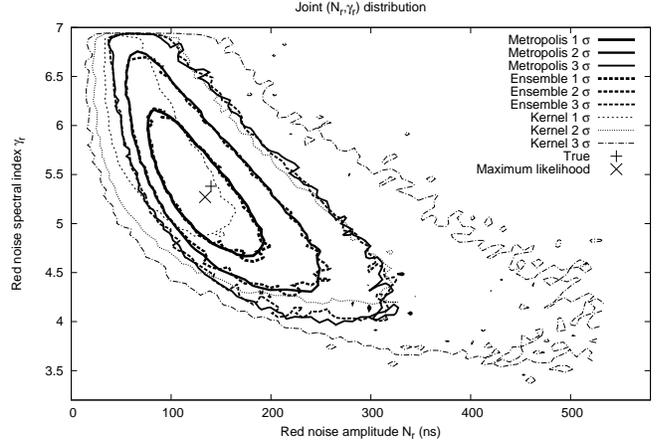}
	\caption{Analysis of the TOAs of
	  Figure~\ref{fig:exampleresidualsensemble} with two methods: a regular
	  MCMC, and the importance sampling method of
	  Equation~(\ref{eq:marginalisedmcmcimportance}). The MCMC contours are
	  marked ``Metropolis'', and the importance sampling contours are
	  marked ``Ensemble''. Also, the contours of the kernel set that has
	  been used are shown, marked by ``Kernel''. In all cased, the
	  $68\%$, $95\%$, and $99.9\%$ contours are shown. The true
	  values for this simulation were: $N_{r}=145$ns and $\gamma_{r}=5.4$.
	  The maximum likelihood values were $N_{r}^{ml}=134$ns and
	  $\gamma_{r}^{ml}=5.27$.}
	\label{fig:ensembleset1combined}
      \end{figure}

\begin{figure}
	\includegraphics[width=0.5\textwidth]{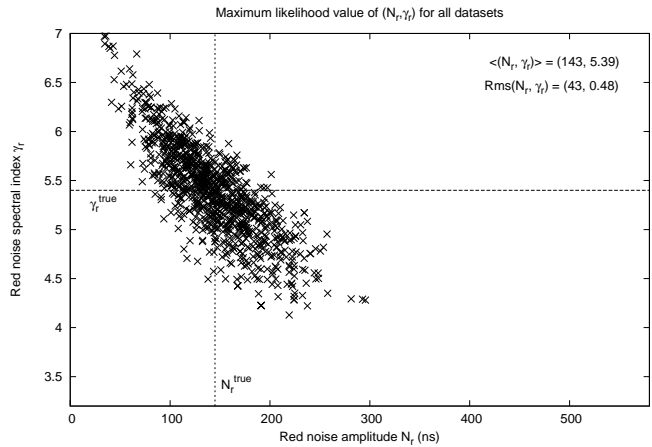}
	\caption{The maximum likelihood values for the parameters $N_{r}$ and
	  $\gamma_{r}$, for the $k=1000$ datasets of
	  Section~\ref{sec:statisticalproperties}. The maximum likelihood values
	  are taken to be the values of the parameters of the MCMC sample with
	  the highest likelihood. This collection of estimators seems to display
	  the same characteristics as the marginalised posterior of
	  Figure~\ref{fig:ensembleset1combined}. As in
	  Figure~\ref{fig:ensembleset1combined}, the true
	  values for this simulation were: $N_{r}=145$ns and $\gamma_{r}=5.4$.
	  The mean values were $\langle (N_{r}, \gamma_{r}) \rangle =
	  (143,5.39)$, and the standard deviations were
	  $\text{Rms}(N_{r},\gamma_{r}) = (43, 0.48)$.}
	\label{fig:mlensemble}
      \end{figure}

      We use the method outlined above to test $k=1000$ datasets. The random
      Gaussian process is a summation of several components:\newline
      1) the error bars of the individual data points of J$1713$+$0747$, as
      described in \citet{vanhaasteren2011}.\newline
      2) an extra component of noise that is added in quadrature to all error
      bars. This parameter is the same for all data points, and represents the
      pulse phase jitter (EQUAD). This random pulse jitter is expected to be one of
      the fundamental limits to pulsar timing precision \citep{Cordes2010}.\newline
      3) a red timing-noise TCSS, described by a power-law spectrum of
      the form $S(f)=N_{r}^{2}(1/1\rm{yr}^{-1})(f/1\rm{yr}^{-1})^{-\gamma_{r}}$,
      with $N_{r}$ the noise amplitude, and $\gamma_{r}$ the spectral index that
      describes the ``redness'' of the timing-noise, with a low-frequency
      cut-off of $f_L = 0.03\text{yr}^{-1}$. As part of our model, we keep
      $f_L$ fixed during our analysis.\newline
      For each dataset, we have three parameters that vary during the MCMC,
      and we have $12$ timing-model parameters that we analytically marginalise
      over during the MCMC. We show the marginalised posterior distribution of
      the red timing noise parameters here as an example in Figure
      \ref{fig:ensembleset1combined}. Our re-weighting
      scheme of Equation~(\ref{eq:marginalisedmcmcimportance}) has modified the
      kernel likelihood correctly to match the true posterior distribution. The
      sample in the MCMC chain that has the highest likelihood value is a 
      good estimator for the maximum likelihood in an MCMC with this few
      dimensions. In Figure~\ref{fig:mlensemble} we present the maximum
      likelihood estimators for the red timing noise parameters that we obtain
      in this way. The collection of estimators of the ensemble seems to
      follow a distribution with the same shape as the marginalised posterior of
      Figure~\ref{fig:ensembleset1combined}, the maximum likelihood value of
      which is also shown to be close to the centre of the distribution of maximum
      likelihood estimators.

      The ensemble analysis has resulted in $k=1000$ distribution functions
      in $15$ dimensions. In order to keep our presentation of the results
      transparent, we restrict our discussion to the $15$k one-dimensional
      marginalised posterior distribution functions that follow from this
      analysis. For each of the $15$k distributions, we have access to the
      true value that we gave as an input to our simulations. A basic check
      would be to verify that for $68\%$ of the datasets, the true value lies
      within the inner $68\%$ of the marginalised posterior distribution.
      We generalise that type of basic check to a more extensive test of both
      the width and the shape of all the $15$k distributions.

\begin{figure}
	\includegraphics[width=0.5\textwidth]{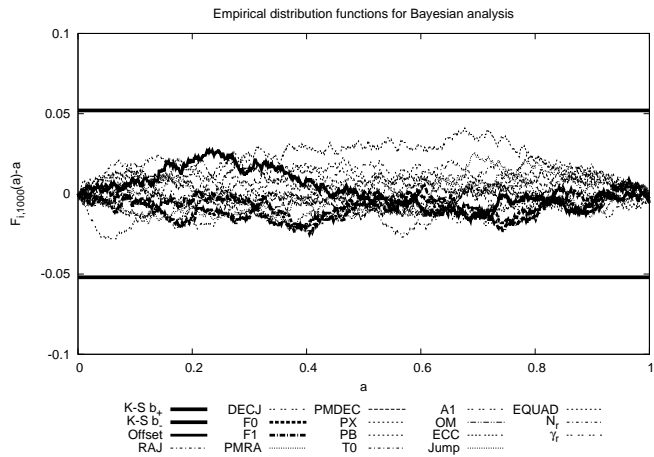}
	\caption{Empirical distribution function $F_{i,k}(a)-a$ for all parameters
	  $i$, with $k=1000$, for the Bayesian analysis. We used mock data of
	  J$1713$+$0747$ of Section~\ref{sec:statisticalproperties}, with red
	  noise modelled with a power-law spectral density. The
	  Kolmogorov-Smirnov boundaries with significance level $\alpha=0.01$
	  are displayed as the $(b_{+},b_{-})$ lines.
	  The {\rm Tempo2} parameter identifiers are:\newline
	  Offset: Unknown absolute phase offset\newline
	  RAJ: Right ascension of the pulsar\newline
	  DECJ: Declination of the pulsar\newline
	  F0: Pulse frequency\newline
	  F1: Pulse frequency derivative\newline
	  PMRAJ: Proper motion in right ascension\newline
	  PMDEC: Proper motion in declination\newline
	  PX: Parallax\newline
	  PB: Orbital period\newline
	  T0: Epoch of periastron\newline
	  A1: Projected semi-major axis of the orbit\newline
	  OM: Longitude of periastron\newline
	  ECC: Eccentricity of the orbit\newline
	  Jump: Random phase jump\newline
	  EQUAD: Random pulse phase jitter}
	\label{fig:edfbayesian}
      \end{figure}

      Provided that our model is correct, the posterior distribution gives the
      probability that the true value of a parameter has a certain value. Since
      we have done many trials, we can count how many times the true value
      $\theta_{i}^{\text{true}}$ of
      parameter $\theta_i$ lies within the most-likely $x\%$ of the posterior
      distribution. By definition of the posterior, for large number of trials
      this number approaches $x\%$ of the total number of trials.
      More formally, we define the inner high-probability
      region (HPR) of the one-dimensional marginalised posterior as:
      \begin{eqnarray}
	\int\limits_{W}\! p\left( \theta_{i}
	\right)\, \mathrm{d}\theta_{i} &=& a, \nonumber \\
	W &=& \left\{\theta_i \in \mathbb{R}: P(\theta_i)>L_{a} \right\},
	\label{eq:innerpercentage}
      \end{eqnarray}
      where $L_{a}$ is some value $>0$ unique for each $a$, where $a$ is a
      probability with $0\leq a\leq 1$. For each parameter,
      we define a threshold value $L_{\text{t}} = P(\theta_i^{\text{true}})$.
      The true value of the parameter lies within the HPR of the 
      marginalised posterior distribution when $L_{\text{t}} > L_a$. By
      definition of the posterior distribution, the probability that the true
      value lies within the HPR is given by ${\rm Pr}(L_{\text{t}} > L_a)=a$,
      where we use ${\rm Pr}$ to denote probabilities.
      We define
      the empirical distribution function (EDF) as \citep{vaart2000}:
      \begin{equation}
	F_{i,k}(a) = \frac{1}{k} \sum\limits_{j=1}^{k}\Theta\left(L_{\text{t}} - L_a
	\right),
	\label{eq:edf}
      \end{equation}
      where $\Theta(x)$ is the Heaviside function, here used as an indicator.
      The term in the summation of Equation~(\ref{eq:edf}) is the indicator for
      the event $L_{\text{t}} > L_a$. For a fixed $L_a$, this is a Bernoulli
      random variable with probability $a$. Hence, $F_{i,k}(a)$ is a binomial
      random variable with mean $a$, and variance $a(1-a)/k$. Therefore, by the
      law of large numbers:
      \begin{equation}
	\lim\limits_{k \to \infty} F_{i,k}\left(a\right) = a.
	\label{eq:limitedf}
      \end{equation}
      The Glivenko-Cantelli theorem \citep{Glivenko1933, Cantelli1933} states
      that this convergence happens uniformly over $a$.
      In Figure \ref{fig:edfbayesian} we present the empirical distribution
      function for all timing-model parameters in our simulation.
      
      We compare our EDF $F_{i,k}(a)$ to the EDF as used in the
      Kolmogorov-Smirnov (K-S) test in statistics, where one can test for the
      equality of a sampled distribution function to a reference distribution
      function.  Given a number of samples, the K-S test statistic quantifies how
      much the distribution of the samples, and the reference distribution are
      alike.  Although we have not one single, but many reference distributions,
      we can define a similar K-S statistic for our EDF:
      \begin{equation}
	D_{i,k} = \sup\limits_{a}\left|F_{i,k}(a)-a\right|.
	\label{eq:ksstatistic}
      \end{equation}
      This K-S statistic is our quantitative test whether or not the
      null-hypothesis -our data analysis method is consistent- should be
      rejected.
      In the canonical application of the K-S
      statistic, one chooses as threshold for the quantity
      $\sqrt{k}D_{i,k}$, which is expected to
      follow a Kolmogorov distribution $P_{k}(K)$:
      \begin{equation}
	\sqrt{k}D_{i,k} > K_{\alpha},
	\label{eq:kstest}
      \end{equation}
      where  our significance $\alpha$ is determined by
      $P_{k}(K \leq K_{\alpha}) = 1 - \alpha$. Two commonly used values are:
      $\alpha=0.05$ with $K_\alpha=1.36$, and $\alpha=0.01$ with
      $K_\alpha=1.63$. We choose our significance level as $\alpha=0.01$, which
      together with $k=1000$ implies that we should reject the null-hypothesis
      that our analysis method is consistent when $D_{i,k} > 0.052$. We can see
      in Figure \ref{fig:edfbayesian} that we do not need to reject the
      null-hypothesis.

  \section{Comparison with the Cholesky method} \label{sec:comparison}
    Recently, CHCMV have proposed a new method to include TCSSs in the analysis
    of pulsar timing observations: the Cholesky method. The Cholesky method
    describes the problem of fitting to the timing model as a whitening problem,
    where both the data and the description of the timing model need to be
    whitened with a
    Cholesky decomposition matrix.  This approach is identical to a GLS fit to the
    timing model given by Equation~(\ref{eq:gls}).  This requires prior
    knowledge of the covariance matrix $C$, which CHCMV substitute with a
    best estimator for the power spectral density of the timing-residuals,
    produced with an advanced spectral analysis method. This spectral analysis
    method is implemented in the form of a {\rm Tempo2} plugin called {\rm
    spectralModel}.
    
    The algorithm implemented in {\rm spectralModel} allows
    determination of the power spectral density of the post-fit
    timing-residuals, assuming that the power spectral density has some specific
    form. For steep red TCSSs as used in this work, the power spectral density
    is modelled as a power-law with a so-called corner frequency $f_c$:
    \begin{equation}
      P\left(f\right) = \frac{A^{2}}{1{\rm yr}^{-1}\left(1+\left(f/f_c\right)^{2}\right)^{\alpha/2}},
      \label{eq:chcmvpower}
    \end{equation}
    where $A$ is the amplitude of the TCSS, $f$ is the frequency, and $\alpha$
    is the spectral index. The difference with a pure power-law as we use in
    Equation~(\ref{eq:powerlawspectraldensity}), is that this power spectral
    density does not diverge for $f \rightarrow 0$. The user needs to provide
    estimates for $\alpha$ and $f_c$ and check that these are correct; fitting
    to the data is only done for the amplitude $A$.

    A direct comparison between the Cholesky method of CHCMV and the
    inference of timing-model parameters with a Bayesian analysis as done in
    this work can be made. Both methods take into account the fact that the
    TOAs may contain a TCSS with a red power spectrum, of which
    estimates can be obtained. And both methods obtain improved estimates of the
    timing-model parameters due to the incorporation of the TCSS contribution to
    the TOAs.  However, several important differences should be
    highlighted\footnote{We emphasise that we only refer to the theoretical
    description of CHCMV. This does not include the practical implementation of
    the Cholesky method, {\rm spectralModel}, which in itself has very useful
    general features such as robust spectral estimation.}.

    Firstly, the modelling of the observations is different. We model
    the TCSS as a stationary random Gaussian process that is added to the
    TOAs, prior to the fitting procedure.
    In the Cholesky method, the TCSS is
    modelled as a stationary signal in the post-fit timing-residuals.
    As we have shown in Figure~\ref{fig:covariance}-\ref{fig:covariancenumall},
    this stationarity breaks down in the fitting process. We believe
    that this raises a question about
    the spectral estimation method of CHCMV, since the post-fit timing-residuals
    cannot be described by a stationary TCSS with a mathematically defined
    spectral density.

    Secondly, CHCMV do not fully account for the covariance between the TCSS and
    the timing-model parameters. The use of an optimal spectral
    estimate in a parameter estimation technique analogous to
    Equation~(\ref{eq:gls}) is not completely appropriate:
    the covariance matrix of the TCSS
    is itself covariant with the timing-model parameters, which results in an
    incorrect covariance matrix for the timing-model parameter estimates, and
    incorrect uncertainties in the spectral estimates. CHCMV show that the
    incorrectness of the uncertainties is significant for the quadratic spindown
    parameters, while it is less of a problem for all the other timing-model
    parameters.

    \begin{figure}
      \includegraphics[width=0.5\textwidth]{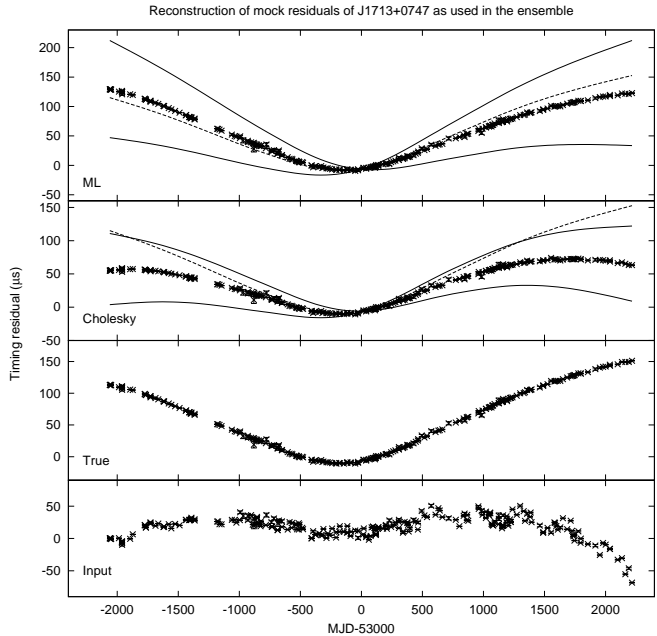}
      \caption{Example of the mock timing-residuals analysed in the ensemble,
	and their reconstruction with various algorithms, all offset from
	each other for clarity. The mock timing
	residuals are based on the observing scheme and timing-model of
	J$1713$+$0747$ as used in
	\citet{vanhaasteren2011}. The error bars in the figure are mostly too
	small to see in this figure, and they vary between different
	observations.
	The TCSS in these residuals comes
	from a source with the following spectral density components:\newline
	1) The error bars of the individual observations.\newline
	2) A white noise component describing the pulse phase jitter (EQUAD), with rms
	$200$ns.\newline
	3) A power-law red noise component
	$S(f)=N_{r}^{2}(1/1\rm{yr}^{-1})(f/1\rm{yr}^{-1})^{-\gamma_{r}}$, with
	amplitude $N_{r}=145$ns, and $\gamma_{r}=5.4$.\newline
	In the figure, four reconstructions of the same realisation are
	shown:\newline
	True: The true timing-residuals as generated by the TCSS.\newline
	ML: The timing-residuals as reconstructed using the maximum likelihood
	values for all parameters: both stochastic parameters and timing-model
	parameters.\newline
	Cholesky: The timing-residuals, reconstructed using a covariance
	matrix produced with
	the {\rm Tempo2} plugin ``{\rm spectralModel}'', which is an
	implementation of the Cholesky method of CHCMV.\newline
	Input: The timing-residuals as produced by {\rm Tempo2} after a normal
	weighted least-squares fit. This ``Input'' set is used as the input
	timing-residuals for all methods.\newline
	In the ML and Cholesky reconstructions, we have marked the true
	timing-residuals as a dashed line, and we have marked the pulse period
	$1$-$\sigma$ boundaries with a solid line. These solid lines demonstrate
	what the residuals would look like if we changed the pulse period, F0,
	by $\pm$ $1$-$\sigma$, and therefore give an impression of how well this
	parameter is determined from the data.}
      \label{fig:exampleresidualsensemble}
    \end{figure}

    In Figure~\ref{fig:exampleresidualsensemble} we present one realisation of
    mock data of the ensemble of datasets we generated for J$1713$+$0747$.
    Besides the true residuals as generated by the random Gaussian process, we
    also present three reconstructions of the timing-residuals:\newline
    1) The input timing-residuals to all analysis methods. These were not the true
    timing-residuals\footnote{We actually worked with TOAs. The residuals
    plotted in Figure~\ref{fig:exampleresidualsensemble} are produced using
    different {\rm Tempo2} ``.par'' files.}, but the timing-residuals after a
    weighted least-squares fit was subtracted from the timing-residuals with
    {\rm Tempo2}\newline
    2) Cholesky timing-residuals. We used the {\rm spectralModel} plugin for
    {\rm Tempo2} to produce an estimate for the covariance matrix of the
    post-fit timing-residuals.  The Cholesky timing-residuals
    are constructed using that estimate and Equation~(\ref{eq:gls}).\newline
    3) ML timing-residuals. We used the maximum likelihood of
    Equation~(\ref{eq:marginalisedlikelihoodnew}) for the stochastic parameters
    $\vec{\phi}$ to produce a best estimator for the covariance matrix $C$. This
    results in the maximum likelihood estimator timing-residuals through
    Equation~(\ref{eq:gls}).

    One can see that the maximum likelihood timing-residuals approximate the
    true timing-residuals slightly better than the Cholesky timing-residuals:
    the Cholesky timing-residuals deviate slightly more at the sides, with the
    true residuals not everywhere inside of the $1$-$\sigma$ bound of the pulse
    frequency, indicating
    an error in the low-frequency behaviour.
    The $1$-$\sigma$ bounds of the pulse frequency (and, not shown in the
    figure, for the pulse frequency derivative) are
    smaller for the Cholesky method than those for the maximum likelihood
    timing-residuals.  These statements were generally true for all the
    realisations of the ensemble of mock datasets.  Besides due to the issues
    raised above, this may also be due to the difference in modelling of the
    power spectral density: the true timing-residuals have been generated with a
    strict power-law, and a (fixed) low-frequency cut-off. However, the results
    here seem to be consistent with Table~$4$ of CHCMV.

    \begin{figure}
      \includegraphics[width=0.5\textwidth]{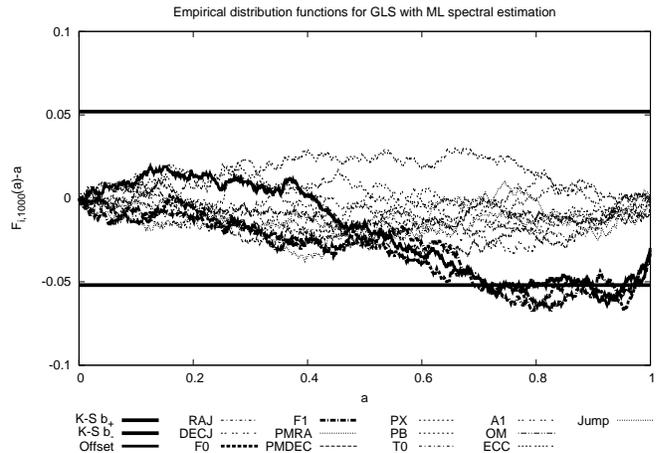}
      \caption{Similar plot as Figure~\ref{fig:edfbayesian}, but now
	for a method that combines a generalised least-squares fit with a
	maximum likelihood spectral estimator (similar to the Cholesky method).
	The same mock datasets as in Figure~\ref{fig:edfbayesian} are used. We
	see that such a method performs well, with results similar to the
	Bayesian analysis. Only the quadratic spindown parameters (the average
	(offset), the pulse period F0, and the period derivative F1) are slightly
	outside the K-S boundaries. This means, assuming Gaussian probability
	distributions, that the rms of the parameter estimates was at least
	$1.11$ times the estimated uncertainty. The parameter labels have the
	same meaning as in Figure~\ref{fig:edfbayesian}.}
      \label{fig:edfml}
    \end{figure}

    We would like to perform a K-S test on the results of the Cholesky method to
    check for consistency. However, this comparison on the ensemble of datasets
    presented in this work would not be fair because the spectral model used by
    the {\rm spectralModel} plugin would then be incorrect. Also, we believe
    that some of the issues with the Cholesky method that we raised above can be
    overcome. We therefore perform a K-S test on the maximum likelihood
    equivalent of the Cholesky method: we use 
    Equation~(\ref{eq:fulllikelihoodnew}) in conjunction with the maximum
    likelihood of Equation~(\ref{eq:marginalisedlikelihoodnew}) as an estimator
    for the covariance matrix $C$. This is equivalent to the Cholesky method
    when using an ``optimal'' estimate for $C$.
    Because this estimator does take into
    account the non-stationary nature of the post-fit residuals, and because the
    modelling is the same as in the marginalised posteriors, this is
    effectively a limit on how well any whitening method can perform.
    Applying this method to the same ensemble of mock data as in
    Section~\ref{sec:statisticalproperties} yields Figure~\ref{fig:edfml}.
    We see that the quadratic spindown parameters are slightly rejected by the
    K-S test, which shows that at least some of the discrepancy found by CHCMV
    for the quadratic spindown parameters is due to 
    due to the covariance of the
    TCSS with the quadratic spindown parameters.
    The width of all the marginalised posterior distributions was similar
    between the approach of Figure~\ref{fig:edfbayesian} and
    Figure~\ref{fig:edfml}, except for the quadratic spindown parameters. For
    the quadratic spindown parameters, the width of the marginalised posterior
    distributions was smaller for the maximum likelihood estimates of
    Figure~\ref{fig:edfml} than for the full Bayesian method of
    Figure~\ref{fig:edfbayesian}.
    
    For the quadratic spindown parameters to be rejected
    by a K-S test of this magnitude means that, assuming Gaussian
    probability distribution functions, the rms of the parameter estimates was
    at least a factor of $1.11$ times larger than the estimated uncertainty.
    Table~$4$ of CHCMV shows that their estimates for the pulse frequency and
    frequency derivative were over a
    factor of three too large, which corresponds to $\sup_{a}|F_{i,k}(a) - a| > 0.43$.
    With the $100$ realisations of mock data they used, the K-S bound
    would be $0.16$. This would be a firm rejection, more so than our maximum
    likelihood estimate. This is at least partially due to the underestimated
    uncertainties for the quadratic spindown parameters. Whether or not there is
    also a bias in the estimates of the Cholesky method for these parameters due
    to incorrect modelling of the covariance function is not clear from the
    current analysis.
    We agree with CHCMV that the Cholesky method
    can be further improved to give more reliable spectral estimates at the very
    low frequencies, and that the Cholesky method performs well for the other
    timing model parameters. One possible way to improve the Cholesky method is
    to use a maximum likelihood estimator for the covariance matrix as we have
    done here, which models the non-stationarity, and which by design does not
    suffer from spectral leakage since it does not rely on a periodogram.

  \section{Conclusions} \label{sec:conclusions}
    We investigate time-correlated stochastic signals (TCSSs) in pulsar timing data
    analysis. TCSSs are significantly influenced by fitting
    procedures that solve for timing-model parameters, and timing-model
    parameter estimates can be biased due to absorption of power the
    TCSS. We formally analyse the covariance between the timing
    model and TCSSs, and obtain closed expressions describing the
    behaviour of the TCSSs when fitting to the timing-model.  New
    results we derive in our analysis:\newline
    1) Proof that the results of the Bayesian analysis are unaffected by use of
    different fitting methods (e.g. (un)weighted least-squares),
    provided that the timing solution has converged.\newline
    2) Closed expressions for the post-fit correlations of signals with known
    power spectra.\newline
    3) Analytical closed expressions for the post-fit covariance function of
    power-law signals with quadratic spindown fitting. This includes proof that
    the low-frequency cut-off is removed up to spectral indices up to
    $\gamma=7$, corresponding to $\alpha=3$ for the GWB.\newline
    4) More computationally efficient expressions for the marginalised posterior
    distribution of vHLML.\newline
    5) An analytical expression of the post-fit rms induced by a stochastic
    gravitational-wave background.\newline
    6) Equations on how to extract the timing-model parameters from Bayesian
    MCMC simulations.\newline
    7) A new method to analyse hundreds of mock datasets simultaneously with a
    Bayesian analysis, without significant computational overhead.\newline
    8) A powerful test to check whether any data analysis method produces
    consistent results, based on the Kolmogorov-Smirnov test.

    We test our method on many realisations of mock data, and find that the
    shape, width, and position of the posterior distributions are consistent
    with the input values of the parameters.  We compare our results to methods
    that use a spectral estimate to whiten the timing-residuals, like
    \citet{Coles2011}, and find that an optimal whitening method performs
    equally well as our own method, except for the quadratic spindown
    parameters, in which case the Bayesian analysis produces more consistent
    results.

  \section*{Acknowledgements}
    This research is supported by the Netherlands organisation for Scientific
    Research (NWO) through VIDI grant 639.042.607. YL's research is supported by
    an Australian Research Council Future Fellowship.

%-- main text ------------------------------------------------------------------

%-- BIBLIOGRAPHY ----------------------------------------------------------------
  \bibliographystyle{mn2e.bst}
%  \bibliography{vanhaasteren}

\begin{thebibliography}{}

\bibitem[\protect\citeauthoryear{{Begelman}, {Blandford} \& {Rees}}{{Begelman}
  et~al.}{1980}]{Begelman1980}
{Begelman} M.~C.,  {Blandford} R.~D.,    {Rees} M.~J.,  1980, Nature, 287, 307

\bibitem[\protect\citeauthoryear{{Blandford}, {Romani} \&
  {Narayan}}{{Blandford} et~al.}{1984}]{Blandford1984}
{Blandford} R.,  {Romani} R.~W.,    {Narayan} R.,  1984, Journal of
  Astrophysics and Astronomy, 5, 369

\bibitem[\protect\citeauthoryear{{Cantelli}}{{Cantelli}}{1933}]{Cantelli1933}
{Cantelli} F.~P.,  1933, Giorn. Ist. Ital. Attuari., 4, 421

\bibitem[\protect\citeauthoryear{{Coles}, {Hobbs}, {Champion}, {Manchester} \&
  {Verbiest}}{{Coles} et~al.}{2011}]{Coles2011}
{Coles} W.,  {Hobbs} G.,  {Champion} D.~J.,  {Manchester} R.~N.,    {Verbiest}
  J.~P.~W.,  2011, \mnras, p.~1523

\bibitem[\protect\citeauthoryear{{Cordes} \& {Shannon}}{{Cordes} \&
  {Shannon}}{2010}]{Cordes2010}
{Cordes} J.~M.,  {Shannon} R.~M.,  2010, ArXiv e-prints

\bibitem[\protect\citeauthoryear{{Demorest}, {Ferdman}, {Gonzalez}, {Nice},
  {Ransom}, {Stairs}, {Arzoumanian} \& {Brazier}}{{Demorest}
  et~al.}{2012}]{Demorest2012}
{Demorest} P.~B.,  {Ferdman} R.~D.,  {Gonzalez} M.~E.,  {Nice} D.,  {Ransom}
  S.,  {Stairs} I.~H.,  {Arzoumanian} Z.,    {Brazier} A.,  2012, ArXiv
  e-prints

\bibitem[\protect\citeauthoryear{{Edwards}, {Hobbs} \& {Manchester}}{{Edwards}
  et~al.}{2006}]{Edwards2006}
{Edwards} R.~T.,  {Hobbs} G.~B.,    {Manchester} R.~N.,  2006, \mnras, 372,
  1549

\bibitem[\protect\citeauthoryear{{Glivenko}}{{Glivenko}}{1933}]{Glivenko1933}
{Glivenko} V.,  1933, Giorn. Ist. Ital. Attuari., 4, 92

\bibitem[\protect\citeauthoryear{{Hobbs}, {Edwards} \& {Manchester}}{{Hobbs}
  et~al.}{2006}]{Hobbs2006}
{Hobbs} G.~B.,  {Edwards} R.~T.,    {Manchester} R.~N.,  2006, \mnras, 369, 655

\bibitem[\protect\citeauthoryear{Jaffe \& Backer}{Jaffe \&
  Backer}{2003}]{Jaffe2003}
Jaffe A.,  Backer D.,  2003, \apj, 583, 616

\bibitem[\protect\citeauthoryear{{Kramer}, {Stairs}, {Manchester},
  {McLaughlin}, {Lyne}, {Ferdman}, {Burgay}, {Lorimer}, {Possenti}, {D'Amico},
  {Sarkissian}, {Hobbs}, {Reynolds}, {Freire} \& {Camilo}}{{Kramer}
  et~al.}{2006}]{Kramer2006}
{Kramer} M.,  {Stairs} I.~H.,  {Manchester} R.~N.,  {McLaughlin} M.~A.,  {Lyne}
  A.~G.,  {Ferdman} R.~D.,  {Burgay} M.,  {Lorimer} D.~R.,  {Possenti} A.,
  {D'Amico} N.,  {Sarkissian} J.~M.,  {Hobbs} G.~B.,  {Reynolds} J.~E.,
  {Freire} P.~C.~C.,    {Camilo} F.,  2006, Science, 314, 97

\bibitem[\protect\citeauthoryear{{Lee}, {Bassa}, {Janssen}, {Karuppusamy},
  {Kramer}, {Smits}, {Stappers}}{{Lee} et~al.}{2012}]{Lee2012}
{Lee} K.~J.,  {Bassa} C.~G.,  {Janssen} G.~H., {Karuppusamy} R., {Kramer} M.,
  {Smits} R., {Stappers} B.~W., 2012, \mnras, 423, 2642


\bibitem[\protect\citeauthoryear{{L{\"o}hmer}, {Lewandowski}, {Wolszczan} \&
  {Wielebinski}}{{L{\"o}hmer} et~al.}{2005}]{Lohmer2005}
{L{\"o}hmer} O.,  {Lewandowski} W.,  {Wolszczan} A.,    {Wielebinski} R.,
  2005, \apj, 621, 388

\bibitem[\protect\citeauthoryear{{Lorimer} \& {Kramer}}{{Lorimer} \&
  {Kramer}}{2005}]{Lorimer2005}
{Lorimer} D.~R.,  {Kramer} M.,  2005, {Handbook of Pulsar Astronomy}

\bibitem[\protect\citeauthoryear{Newman \& Barkema}{Newman \&
  Barkema}{1999}]{Newman1999}
Newman M.,  Barkema G.,  1999, Monte Carlo Methods in Statistical Physics.
Oxford University Press Inc., pp 31--86

\bibitem[\protect\citeauthoryear{{Phinney}}{{Phinney}}{2001}]{Phinney2001}
{Phinney} E.~S.,  2001, ArXiv Astrophysics e-prints

\bibitem[\protect\citeauthoryear{Press, Teukolsky, Vetterling \&
  Flannery}{Press et~al.}{1992}]{press1992}
Press W.,  Teukolsky S.,  Vetterling W.,    Flannery B.,  1992, {Numerical
  Recipes in C}, 2nd edn.
Cambridge University Press, Cambridge, UK

\bibitem[\protect\citeauthoryear{{Sesana}, {Vecchio} \& {Colacino}}{{Sesana}
  et~al.}{2008}]{Sesana2008}
{Sesana} A.,  {Vecchio} A.,    {Colacino} C.~N.,  2008, \mnras, 390, 192

\bibitem[\protect\citeauthoryear{{Shannon} \& {Cordes}}{{Shannon} \&
  {Cordes}}{2010}]{Shannon2010}
{Shannon} R.~M.,  {Cordes} J.~M.,  2010, \apj, 725, 1607

\bibitem[\protect\citeauthoryear{{Taylor} \& {Weisberg}}{{Taylor} \&
  {Weisberg}}{1982}]{Taylor1982}
{Taylor} J.~H.,  {Weisberg} J.~M.,  1982, \apj, 253, 908

\bibitem[\protect\citeauthoryear{Vaart}{Vaart}{2000}]{vaart2000}
Vaart A.,  2000, Asymptotic statistics.
Cambridge series on statistical and probabilistic mathematics, Cambridge
  University Press

\bibitem[\protect\citeauthoryear{{van Haasteren} \& {Levin}}{{van Haasteren} \&
  {Levin}}{2010}]{vanhaasteren2010}
{van Haasteren} R.,  {Levin} Y.,  2010, \mnras, 401, 2372

\bibitem[\protect\citeauthoryear{{van Haasteren}, {Levin}, {Janssen},
  {Lazaridis}, {Kramer}, {Stappers}, {Desvignes}, {Purver} \& {Lyne}}{{van
  Haasteren} et~al.}{2011}]{vanhaasteren2011}
{van Haasteren} R.,  {Levin} Y.,  {Janssen} G.~H.,  {Lazaridis} K.,  {Kramer}
  M.,  {Stappers} B.~W.,  {Desvignes} G.,  {Purver} M.~B.,    {Lyne} A.~G.,
  2011, \mnras, 414, 3117

\bibitem[\protect\citeauthoryear{{van Haasteren}, {Levin}, {McDonald} \&
  {Lu}}{{van Haasteren} et~al.}{2009}]{vanhaasteren2009}
{van Haasteren} R.,  {Levin} Y.,  {McDonald} P.,    {Lu} T.,  2009, \mnras,
  395, 1005

\bibitem[\protect\citeauthoryear{Wyithe \& Loeb}{Wyithe \&
  Loeb}{2003}]{Wyithe2003}
Wyithe J.,  Loeb A.,  2003, \apj, 595, 614

\end{thebibliography}

%-- bibliography ----------------------------------------------------------------

\appendix

  \section{Power-law covariance functions} \label{sec:appendixa}
    In this Appendix, we analytically derive the post-fit
    covariance function $WC^{\rm PL}W^{T}$ from
    Equation~(\ref{eq:projectedpostfitcovariance}) of the main text. We rewrite the relevant expressions
    for the basis-functions and the projection operators here for convenience,
    with the same notation as in Section \ref{sec:analyticalcovariance}:
    \begin{eqnarray}
      \label{sec:ap:rewrite}
      \left\langle\vec{x}, \vec{y}\right\rangle_{E} &\approx& \frac{1}{\sigma^{2}
      \Delta t}\int_{-T}^{T}\! x(t)y(t) \, \rm{d}t \\
      \hat{f}_1(t) &=& \frac{1}{\sqrt{2}}\sigma\sqrt{\frac{\Delta t}{T}}
      \nonumber \\
      \hat{f}_2(t) &=& \sqrt{\frac{3}{2}}\sigma\sqrt{\frac{\Delta
      t}{T}}\frac{t}{T} \nonumber \\
      \hat{f}_2(t) &=& \sqrt{\frac{45}{8}}\sigma\sqrt{\frac{\Delta
      t}{T}}\left[\left(\frac{t}{T}\right)^{2}-\frac{1}{3}\right]. \nonumber \\
      C^{\rm P}(t_0,t_3) &=& S(t_0, t_1)C^{\rm PL}(t_1,t_2)S(t_2, t_3) \nonumber \\
      S(t_k, t_l) &=& \sigma^2\Delta t \delta\left(t_k-t_l\right) -
      \sum_{i=1}^{3}
      \hat{f}_{i}(t_k)\hat{f}_{i}(t_l),
      \nonumber
    \end{eqnarray}
    hereafter we always sum over the repeated indices
    $t_1$ and $t_2$.  Because the pre-fit covariance function of a power-law
    spectral density depends only on $\tau=2\pi|t_0-t_3|$, we first
    calculate the following quantity:
    \begin{equation}
      Z^{\rm P}_{\zeta}\left(t_0, t_3\right) = S(t_0,
      t_1)\left|t_1-t_2\right|^{\zeta}S(t_2, t_3).
      \label{eq:covariancetauapp}
    \end{equation}
    We can then construct $C^{\rm P}$ with several $Z^{\rm
    P}_{\zeta}$ terms\footnote{We note that in general $c_{ij}=|t_i-t_j|^{\zeta}$ is
    not a PDS matrix, and it therefore does not correspond to a physical
    stochastic process.}. We write the resulting $Z^{\rm P}_{\zeta}$ in the
    following terms:
    \begin{equation}
      Z^{\rm P}_{\zeta}\left(t_0, t_3\right) = \left|t_0-t_3\right|^{\zeta} -
      \sum_{i=1}^{3}Z_{i}\left(t_0, t_3\right) +
      \sum_{i,j=1}^{3}Z_{ij}\left(t_0,t_3\right).
      \label{eq:tzeta}
    \end{equation}
    The $Z_{ij}$ terms are symmetric in $i$ and $j$, and after evaluation of the
    (somewhat tedious) integrals we find\footnote{The calculations are available
    from the authors by request.}:
    \begin{eqnarray}
      Z_{ij}\left(t_0,t_3\right) &=& \left[\hat{f}_{i}\left(t_1\right)
	\left|t_1-t_2\right|^{\zeta}\hat{f}_{j}\left(t_2\right)\right]
	\hat{f}_{i}\left(t_0\right)
	\hat{f}_{j}\left(t_3\right) \nonumber \\
      U_{11}\left(t_0,t_3\right) &=& \frac{1}{2(1+\zeta)(2+\zeta)}\nonumber \\
      U_{12}\left(t_0,t_3\right) &=& 0 \nonumber \\
      \label{eq:tij}
      U_{13}\left(t_0,t_3\right) &=&
      \frac{15\zeta\left(\left(\frac{t_0}{T}\right)^{2}+\left(\frac{t_3}{T}\right)^{2}-\frac{2}{3}\right)}{4(2+\zeta)(3+\zeta)(4+\zeta)} \\
      U_{22}\left(t_0,t_3\right) &=& -\frac{9\zeta\frac{t_{0}t_{3}}{T^2}}{2(1+\zeta)(2+\zeta)(4+\zeta)}\nonumber \\
      U_{23}\left(t_0,t_3\right) &=& 0 \nonumber \\
      U_{33}\left(t_0,t_3\right) &=&
      \frac{225\zeta(\zeta-2)\left(\left(\frac{t_0}{T}\right)^2-\frac{1}{3}\right)\left(\left(\frac{t_3}{T}\right)^2-\frac{1}{3}\right)}{8(1+\zeta)(2+\zeta)(4+\zeta)(6+\zeta)}\nonumber
    \end{eqnarray}
    where:
    \begin{equation}
      U_{ij} = \left\{\begin{array}{ll}
	\frac{T^2}{(2T)^{2+\zeta}}\left(Z_{ij} + Z_{ji}\right) & \textrm{if $i\neq j$} \\
	\frac{T^2}{(2T)^{2+\zeta}}Z_{ij} & \textrm{if $i = j$}
      \end{array} \right. .
      \label{eq:uij}
    \end{equation}
    We find for the $Z_{i}$ terms:
    \begin{eqnarray}
      \label{eq:ti}
      Z_{i}\left(t_0,t_3\right) &=&
	\hat{f}_{i}\left(t_0\right)
        \hat{f}_{i}\left(t_1\right)
	\left|t_1-t_3\right|^{\zeta} \\
      &+& \left|t_0-t_2\right|^{\zeta}
	\hat{f}_{i}\left(t_2\right)
        \hat{f}_{i}\left(t_3\right) \nonumber \\
      Z_{1}\left(t_0,t_3\right) &=& \frac{\left(T+t_0\right)^{\zeta+1}+\left(T-t_0\right)^{\zeta+1}}
	{2T(1+\zeta)} \nonumber \\
      &+& \frac{\left(T+t_3\right)^{\zeta+1}+\left(T-t_3\right)^{\zeta+1}}
	{2T(1+\zeta)} \nonumber \\
      Z_{2}\left(t_0,t_3\right) &=& \frac{3\left(-\frac{t_3}{T}\left(T+t_0\right)^{\zeta+1}+
        \frac{t_3}{T}\left(T-t_0\right)^{\zeta+1}\right)}
	{2T(1+\zeta)} \nonumber \\
      &+& \frac{3\left(-\frac{t_0}{T}\left(T+t_3\right)^{\zeta+1}+
        \frac{t_0}{T}\left(T-t_3\right)^{\zeta+1}\right)}
	{2T(1+\zeta)} \nonumber \\
      &+& \frac{3\left(\frac{t_0}{T}\left(T+t_3\right)^{\zeta+2}
        -\frac{t_0}{T}\left(T-t_3\right)^{\zeta+2}\right)}
	{2T^{2}(1+\zeta)(2+\zeta)} \nonumber \\
      &+& \frac{3\left(\frac{t_3}{T}\left(T+t_0\right)^{\zeta+2}
        -\frac{t_3}{T}\left(T-t_0\right)^{\zeta+2}\right)}
	{2T^{2}(1+\zeta)(2+\zeta)} \nonumber \\
      Z_{3}\left(t_0,t_3\right) &=& \frac{15\left(\left(\frac{t_0}{T}\right)^{2}-\frac{1}{3}\right)\left(
	\left(T+t_3\right)^{\zeta+1}+\left(T-t_3\right)^{\zeta+1}\right)}{4T(1+\zeta)}
	\nonumber \\
      &+& \frac{15\left(\left(\frac{t_3}{T}\right)^{2}-\frac{1}{3}\right)\left(
	\left(T+t_0\right)^{\zeta+1}+\left(T-t_0\right)^{\zeta+1}\right)}{4T(1+\zeta)}
	\nonumber \\
      &-& \frac{45\left(\left(\frac{t_0}{T}\right)^{2}-\frac{1}{3}\right)\left(
	\left(T+t_3\right)^{\zeta+2}+\left(T-t_3\right)^{\zeta+2}\right)}{4T^{2}(1+\zeta)(2+\zeta)}
	\nonumber \\
      &-& \frac{45\left(\left(\frac{t_3}{T}\right)^{2}-\frac{1}{3}\right)\left(
	\left(T+t_0\right)^{\zeta+2}+\left(T-t_0\right)^{\zeta+2}\right)}{4T^{2}(1+\zeta)(2+\zeta)}
	\nonumber \\
      &+& \frac{45\left(\left(\frac{t_0}{T}\right)^{2}-\frac{1}{3}\right)\left(
	\left(T+t_3\right)^{\zeta+3}+\left(T-t_3\right)^{\zeta+3}\right)}{4T^{3}(1+\zeta)(2+\zeta)(3+\zeta)}
	\nonumber \\
      &+& \frac{45\left(\left(\frac{t_3}{T}\right)^{2}-\frac{1}{3}\right)\left(
	\left(T+t_0\right)^{\zeta+3}+\left(T-t_0\right)^{\zeta+3}\right)}{4T^{3}(1+\zeta)(2+\zeta)(3+\zeta)}.
	\nonumber
    \end{eqnarray}
    Then $C^{\rm P}$ are obtained by substituting the above expressions into
    the following:
    \begin{eqnarray}
      C^{\rm P} &=& 
	A^2\left(\frac{1\rm{yr}^{-1}}{f_{L}}\right)^{\gamma-1}
	  \left\{ \Gamma(1-\gamma)\sin\left(\frac{\pi\gamma}{2}\right)
	  \left(f_{L} 2\pi\right)^{\gamma-1} Z^{\rm P}_{\gamma-1} \right.\nonumber\\
	& &-\left.\sum_{n=0}^{\infty}\left(-1\right)^{n}
	  \frac{\left(f_{L}2\pi\right)^{2n}}{(2n)!
	  \left(2n+1-\gamma\right)}Z^{\rm P}_{2n}\right\} . 
      \label{eq:thewholeshebang}
    \end{eqnarray}
    Interestingly, $Z^{\rm P}_{0} = Z^{\rm
    P}_{2} = Z^{\rm P}_{4} = 0$. This means that for $\gamma<7$, all the $f_L$
    dependent terms in $C^{\rm P}$ vanish due to the removal of quadratic
    spindown.

    In the calculations of the rms of signals, we also need the following
    integral, valid for $\zeta>0$:
    \begin{equation}
      \frac{1}{2T}\int_{-T}^{T} \! Z^{\rm P}_{\zeta}\left(t,
      t\right) \, \rm{d}t
      =
      \frac{3(4-\zeta)(\zeta-2)2^{1+\zeta}T^{\zeta}}{(1+\zeta)(2+\zeta)(4+\zeta)(6+\zeta)}
      \label{eq:zPzetaint}
    \end{equation}
    This result does not contradict $Z^{\rm P}_{0} = 0$, since for $\zeta = 0$
    the integral does not exist: for $\gamma \leq 1$ we also need a
    high-frequency cut-off for the power spectral density.

  \section{Gaussian priors and timing model analysis} \label{sec:appendixb}
    In this Appendix we show how to include Gaussian priors for the timing
    model parameters efficiently. We also present a derivation of
    Equation~(\ref{eq:marginalisedfullposterior}).

    \subsection{Gaussian priors}
      Besides with flat priors, analytically marginalising over timing-model
      parameters is also possible with Gaussian priors for the the timing-model
      parameters.  We define Gaussian priors for the $m$ timing-model parameters
      $\vec{\xi}$ as:
      \begin{equation}
	P_{0}\left( \vec{\xi} \right) =
	\frac{
	  \exp \left( \frac{-1}{2}\left(
	  \vec{\xi}-\vec{\xi}_0 \right)^{T}\Sigma_{0}^{-1}\left(
          \vec{\xi}-\vec{\xi}_0 \right)\right)
	}{
	  \sqrt{(2\pi)^{m}\det\Sigma_{0}}
	},
	\label{eq:gaussianprior}
      \end{equation}
      where $\vec{\xi}_{0}$ are the maxima of the prior probabilities, and
      $\Sigma_{0}$ is the $(m\times m)$ prior covariance matrix of the timing
      model parameters. We now proceed with Equation~(\ref{eq:likelihood})
      multiplied with this prior, and rewrite this analogous to what we did in
      Equation~(\ref{eq:fulllikelihoodnew}):
      \begin{eqnarray}
	P\left( \vec{\phi}, \vec{\xi} | \vec{\delta t} \right) &=& \frac{
	  \exp\left(\frac{-1}{2}\left[ \vec{\delta t}^{T}C^{-1}\vec{\delta t}+
	    \vec{\chi}^{T}\Sigma^{-1}\vec{\chi}+
	    \vec{\xi}_{0}^{T}\Sigma_{0}^{-1}\vec{\xi}_{0} \right]\right)
	}{
	  \sqrt{\left( 2\pi \right)^{n+m}\det\Sigma_0\det C}
	} \nonumber \\
	&\times& \exp\left( \frac{-1}{2}\left( \vec{\xi}-\vec{\chi} \right)^{T} 
	\Sigma^{-1}\left( \vec{\xi}-\vec{\chi} \right)^{T}\right),
	\label{eq:posteriorgaussianprior}
      \end{eqnarray}
      with
      \begin{eqnarray}
	\vec{\chi} &=& \left( M^{T}C^{-1}M+\Sigma_{0}^{-1} \right)^{-1}\left(
	M^{T}C^{-1}\vec{\delta t}+\Sigma_{0}^{-1}\vec{\xi}_{0}
	\right) \nonumber \\
	\Sigma^{-1} &=& M^{T}C^{-1}M + \Sigma_{0}^{-1}.
	\label{eq:gaussiandefs}
      \end{eqnarray}
      Up to a normalisation constant due to the inclusion of the prior, these
      expressions reduce to Equation~(\ref{eq:fulllikelihoodnew}) if we take
      $\Sigma^{-1}_{0} = 0$, and $\vec{\xi}_{0}=0$. Marginalising
      Equation~(\ref{eq:posteriorgaussianprior}) over the timing-model
      parameters gives:
      \begin{eqnarray}
	\label{eq:marginalisedposteriorgaussian}
	P\left( \vec{\phi} | \vec{\delta t} \right) &=& \frac{
	  \sqrt{\det \Sigma}
	}{
	  \sqrt{\left( 2\pi \right)^{n}\det\Sigma_0\det C}
	} \\
	&\times&
	\exp\left(\frac{-1}{2}\left[ \vec{\delta t}^{T}C^{-1}\vec{\delta t}+
	    \vec{\chi}^{T}\Sigma^{-1}\vec{\chi}+
	    \vec{\xi}_{0}^{T}\Sigma_{0}^{-1}\vec{\xi}_{0} \right]\right),
	    \nonumber
      \end{eqnarray}

    \subsection{Posteriors for the timing-model parameters}
      The MCMC samples are drawn from $P( \vec{\phi} | \vec{\delta t})$ of
      Equation~(\ref{eq:marginalisedposteriorgaussian}), which is $P(
      \vec{\phi}, \vec{\xi} | \vec{\delta t} )$ of
      Equation~(\ref{eq:posteriorgaussianprior}) marginalised over $\vec{\xi}$.
      However, we are interested in
      interested in $P(\vec{\xi} | \vec{\delta t} )$, which is $P(
      \vec{\phi}, \vec{\xi} | \vec{\delta t} )$ marginalised over all stochastic
      parameters $\vec{\phi}$.
      Using an importance sampling approach, we
      approximate the full posterior distribution with the MCMC samples as:
      \begin{eqnarray}
	\label{eq:approxfullposteriorgaussian}
	P\left( \vec{\xi} | \vec{\delta t} \right) &\approx& \left\langle
	\frac{P\left( \vec{\phi}, \vec{\xi} | \vec{\delta t} \right)}{P\left(
	\vec{\phi} | \vec{\delta t}\right)}
	\right\rangle \\
	&=& \left\langle\frac{\exp\left( \frac{-1}{2}\left( \vec{\xi}-\vec{\chi}
	\right)^{T} \Sigma^{-1}\left( \vec{\xi}-\vec{\chi} \right)^{T}\right)
	}{\sqrt{\left( 2\pi \right)^{m}\det \Sigma}
	}\right\rangle, \nonumber
      \end{eqnarray}
      where we use $\langle\dots\rangle$ to average over all MCMC samples.

      We assume that we would like to obtain the $2$-dimensional marginalised
      posterior as a function of the parameters $\xi_k$ and $\xi_l$, with $1
      \leq k,l \leq m$, but the generalisation to a different dimensionality is
      straightforward. The $2$-dimensional marginalised posterior is constructed
      by integrating over all elements of $\xi$, except for $\xi_k$ and $\xi_l$.
      This integration is analogous to what we did with
      Equation~(\ref{eq:likelihood})-(\ref{eq:marginalisedlikelihoodnew}). We
      therefore construct two auxiliary matrices similar to $F$ and $G$ of
      Equation~(\ref{eq:orthonormalbases}):
      \begin{eqnarray}
	L_F &=&
	\begin{pmatrix}
	  \cdots & \hat{e_{k-1}} & \hat{e_{k+1}} & \cdots
	  &
	  \hat{e_{l-1}} & \hat{e_{l+1}} & \cdots
	\end{pmatrix} \nonumber \\
	L_G &=& \begin{pmatrix}
	  \hat{e_{k}} & \hat{e_{l}}
	\end{pmatrix},
	\label{eq:designmatrixlflg}
      \end{eqnarray}
      where the $\hat{e_i}$ are the basis vectors of ${\mathbb R}^m$. Similar to
      Equation~(\ref{eq:marginalisedlikelihoodnew}), the $2$-dimensional marginalised
      posterior now becomes:
      \begin{equation}
	P\left(\xi_k, \xi_l | \vec{\delta t} \right) = \left\langle
	\frac{\exp\left(\frac{-1}{2}\vec{\Delta
	\xi}^{T}L_{G}\left(L_{G}^{T}\Sigma L_{G}\right)^{-1}L_{G}^{T}\vec{\Delta
	\xi}\right)
	}{\sqrt{\left(2\pi\right)^{2}\det
	\left(L_{G}^{T}\Sigma L_{G}\right)}
	}\right\rangle
	\label{eq:marginalisedfullposteriora}
      \end{equation}
      where :
      \begin{equation}
	\vec{\Delta \xi} =
	\begin{pmatrix}
	  \xi_{k} - \chi_{k} \\
	  \xi_{l} - \chi_{l}
	\end{pmatrix}.
	\label{eq:DeltaXib}
      \end{equation}

\end{document}